\documentclass[aps,prd,10pt,showpacs,amsmath,showkeys,twocolumn,floatfix,amssymb, preprintnumbers, nofootinbib, superscriptaddress]{revtex4}
\usepackage{graphicx}
\usepackage{comment}
\usepackage[usenames]{color}
\usepackage{bm}
\usepackage{ifpdf}
\usepackage{floatrow}
\usepackage{makecell}
\usepackage{stackrel}
\usepackage{subcaption}
\usepackage{enumitem}

\usepackage[normalem]{ulem}
\usepackage[dvipsnames]{xcolor}
\usepackage[utf8]{inputenc}
\usepackage{hyperref}
\hypersetup{
  pdfnewwindow=true,      % links in new window
  colorlinks=true,        % false: boxed links; true: colored links
  linkcolor=PineGreen,    % color of internal links
  citecolor=PineGreen,    % color of links to bibliography
  filecolor=PineGreen,    % color of file links
  urlcolor=PineGreen      % color of external links
}
\newcommand{\be}{\begin{eqnarray}}
\newcommand{\ee}{\end{eqnarray}}

\begin{document}

\title{Toward extracting the scattering phase shift from integrated correlation functions II:  a relativistic lattice field theory model}

\author{Peng~Guo}
\email{peng.guo@dsu.edu}

\affiliation{College of Arts and Sciences,  Dakota State University, Madison, SD 57042, USA}
\affiliation{Kavli Institute for Theoretical Physics, University of California, Santa Barbara, CA 93106, USA}

\date{\today}

\begin{abstract}
 In present work, a relativistic relation that connects the difference of interacting and non-interacting integrated two-particle correlation functions in finite volume to infinite volume scattering phase shift through an integral  is derived. We show that the difference of integrated finite volume correlation functions converge rapidly to its infinite volume limit as the size of periodic box is increased.   The fast convergence  of our proposed formalism is  illustrated  by   analytic solutions of a contact interaction model,  the perturbation theory calculation, and  also Monte Carlo simulation of a complex $\phi^4$ lattice field theory model.
\end{abstract}

\maketitle

\section{Introduction}\label{sec:intro}

   Few nucleons interactions provide crucial inputs to  nuclear many-body studies of matter:  such as  the neutron-star equation of state and   stability of neutron-rich isotopes \cite{doi:10.1146/annurev-nucl-102313-025446},     exotic decays of various nuclei \cite{Navrátil_2016},    and  experimental searches for new physics beyond-Standard-Model particles  \cite{Engel_2017}.  The fundamental theory of nuclear physics is quantum chromodynamics (QCD), the theory of the strong interaction between quarks mediated by gluons.    The only model-independent and systematically improvable method for computing the properties and interactions of nucleons directly from QCD is  lattice QCD (LQCD), which is  the Euclidean spacetime formulation of QCD on a finite and discrete lattice in a periodic hypercubic box.       As the consequence of calculation in Euclidean spacetime and in finite box,   the energy spectrum become discrete, no  asymptotic states and no direct access to scattering amplitudes are available,  and finite volume effect must be taken into account.

To map out  nucleon-nucleon scattering amplitudes in LQCD calculation,  L\"uscher formula \cite{Luscher:1990ux}  has been widely used, which relates discrete energy levels of finite volume systems to      their scattering phase shifts in a compact form.  The typical two-step procedure are:   (1) firstly  extracting   low-lying   energy spectrum   by fitting exponential decaying behavior of correlation functions in Euclidean spacetime, and looking for the plateau in temporal correlation functions when Euclidean time is large enough so that   the lowest energy level becomes dominant and correlation functions are free of  excited states pollution;  (2)       the discrete energy spectra   thus are converted into scattering phase shifts  by applying L\"uscher formula.   The two-step  L\"uscher formula approach has been proven very successful in number of applications especially in meson sector, see e.g. Refs.~\cite{Aoki:2007rd,Feng:2010es,Lang:2011mn,Aoki:2011yj,Dudek:2012gj,Dudek:2012xn,Wilson:2014cna,Wilson:2015dqa,Dudek:2016cru,Beane:2007es, Detmold:2008fn, Horz:2019rrn,Guo:2020kph}.  The  formalism  has been quickly extended to include inelastic effects, such as coupled-channel effect and three-body problems,  etc.  see e.g. Refs.~\cite{Rummukainen:1995vs,Christ:2005gi,Bernard:2008ax,He:2005ey,Lage:2009zv,Doring:2011vk,Guo:2012hv,Guo:2013vsa,Kreuzer:2008bi,Polejaeva:2012ut,Hansen:2014eka,Mai:2017bge,Mai:2018djl,Doring:2018xxx,Guo:2016fgl,Guo:2017ism,Guo:2017crd,Guo:2018xbv,Mai:2019fba,Guo:2018ibd,Guo:2019hih,Guo:2019ogp,Guo:2020wbl,Guo:2020kph,Guo:2020iep,Guo:2020ikh,Guo:2020spn,Guo:2021lhz,Guo:2021uig,Guo:2021qfu,Guo:2021hrf}.    
 Unfortunately the application of  two-step L\"uscher formula approach in two-nucleon systems is hindered by a few challenges:

 \begin{itemize}

\item     The signal-to-noise ratio (S/N) \cite{lepage1989analysis,DRISCHLER2021103888} in stochastic evaluation of the path integral for the correlation of two-nucleon at large Euclidean times behave as
\begin{equation}
\mathcal{R}(\tau)  \stackrel{\tau \rightarrow \infty}{\sim }  e^{-  (m_N - \frac{3}{2} m_{\pi}) \tau},
\end{equation}
where $m_N$ and $m_{\pi}$ are the nucleon and pion mass.  Exponentially more statistics are   required to overcome   S/N problem.

\item  Large volume  leads to significant increase of   density of states with  small splitting between energy levels: $\bigtriangleup E \approx \frac{1}{m_N} (\frac{2\pi}{L} )^2$, where $L$ is a spatial extension of the lattice.  The required Euclidean time $\tau$ to display the signal of clear plateau must be $\tau \gg (\bigtriangleup E)^{-1} \sim m_N (L/2\pi)^2$, which could be  well into the region where the noise has swamped the signal.

\item  The difficulties of  L\"uscher formula approach at large volume limit  increase significantly   due to  the increasingly dense finite-volume spectrum and the large number of  interpolating operators that are required to faithfully project out desired low-lying energy levels, see e.g. Ref.~\cite{Bulava:2019kbi}.

\end{itemize}
 The challenges have prevented substantive progress on calculation of two-nucleon interactions in LQCD. Even with all the technological advancements in computational science, there are still no calculation of two-nucleon systems utilizing the L\"uscher formula with pion masses lighter than $300 \mbox{ MeV}$.

  These challenges also motivate exploration of alternatives to the two-step L\"uscher  formula approach. The HAL QCD collaboration potential method \cite{PhysRevLett.99.022001,10.1143/PTP.123.89,PhysRevD.99.014514,ISHII2012437,AokiEPJA2013}   was developed a decade ago and may offer an alternative approach to determining two-nucleon interactions at low-energies without ground state saturation. Unfortunately, results from the two-step L\"uscher  formula method and HAL QCD collaboration potential method do not agree, not even qualitatively at very heavy pion mass \cite{DRISCHLER2021103888}.  The discrepancy between L\"uscher  formula approach and HAL QCD collaboration potential method  is known as the two-nucleon controversy and poses a severe challenge to LQCD prediction on few-nucleon dynamics.  Some other new  ideas have also been proposed in recent years, such as, determining scattering amplitudes from finite-volume spectral functions in  Ref.~\cite{Bulava:2019kbi},   extraction of spectral densities from lattice correlators in Refs.~\cite{Hansen:2019idp,Bailas:2020qmv}, and extracting phase shifts from integrated correlation functions \cite{Guo:2023ecc}.

In Ref.~\cite{Guo:2023ecc},  we show that the difference of  integrated  finite volume  two-particle correlation functions  between interacting   and free  nonrelativistic  particles system to infinite volume is related to  scattering phase shift, $\delta(\epsilon)$,  through a integral weighted by a factor $e^{- \epsilon \tau}$,
\begin{equation}
  C  (t)  - C_0 (t)     \stackrel[t= - i \tau]{L \rightarrow \infty}{\rightarrow} \frac{\tau}{\pi}   \int_0^\infty   d  \epsilon  \delta (\epsilon)      e^{-    \epsilon  \tau}   ,   \label{prevresult}
\end{equation}
where $C(t)$ and $C_0(t)$ are integrated correlation functions for two nonrelativistic   interacting and noninteracting particles in the finite box respectively.  $L$ stands for size of periodic box, and   $t $ and $\tau$ are Minkowski and Euclidean time respectively.  Most importantly, we also demonstrated in Ref.~\cite{Guo:2023ecc} that  the  difference of  integrated  finite volume  two-particle correlation functions  rapidly approaches its infinite volume limit that is given by right-hand-side of Eq.(\ref{prevresult}) at short Euclidean time,  $\tau \ll L$,  even with a modest small size box. The fast convergence feature of  Eq.(\ref{prevresult}) at short Euclidean time ($\tau/L \ll 1$) make it  potentially  a good candidate to overcome S/N problem in two-nucleon LQCD calculation.  The proposal of Ref.~\cite{Guo:2023ecc} in principle is   free from issues, such as increasingly dense energy spectra at large volume and ground state saturation, etc. Hence,  it also offer a more suitable framework to overcome the challenges that conventional  two-step Lu\"scher formula   faces  at large volume limit.

   The aim of present work is to extend the nonrelativistic formalism proposed in  Ref.~\cite{Guo:2023ecc} to  a relativistic one. After installing all the relativistic kinematic factors, we will show later on that the relativistic version of  Eq.(\ref{prevresult}) in $1+1$ dimensions is   given by
   \begin{align}
 \triangle C  (t)  &  \stackrel{t = - i \tau}{ =} \sum_{n=0}^\infty \left [ \frac{e^{- E_n \tau}}{E_n} - \frac{e^{- E^{(0)}_n \tau}}{E^{(0)}_n}  \right ]    \nonumber \\
&   \stackrel[t= - i \tau]{L \rightarrow \infty}{\rightarrow} \frac{1}{\pi}   \int_{2m}^\infty   d  \epsilon  \delta (\epsilon)  \left ( \tau+ \frac{1}{\epsilon} \right )   \frac{   e^{-    \epsilon  \tau}  }{\epsilon}  ,   \label{mainresult}
\end{align}
where $E_n$ and $E_n^{(0)}$ are eigen-energy of interacting and noninteracting  relativistic two particles system respectively, and $m$ stands for the mass of two identical particles. The fast convergence of  $ \triangle C  (t) $ into its infinite volume limit given by relation in Eq.(\ref{mainresult}) will be illustrated by (1) an exactly solvable contact interaction model, (2) perturbation theory calculation, and (3) Monte Carlo simulation of a complex $\phi^4$ lattice field theory model.

The paper is organized as follows. First of all,  a field theory model for the study of relativistic spinless particles interaction  is set up in Sec.~\ref{modelsetup}.   The derivation of the infinite volume limit of integrated two-particle correlation function,   its relation to particles scattering phase shift, and exact solutions and perturbation calculation of contact interaction results are all presented in  Sec.~\ref{modelsetup}.     The 1D Monte Carlo simulation test  of $\phi^4$ field  is presented and discussed in  Sec.~\ref{MCsimulation}.  The discussions and summary are given in Sec.~\ref{summary}.

\section{A lattice field theory model}\label{modelsetup}    
In present work, we will consider a relativistic lattice field theory  model for the interaction of charged scalar particles via a short range potential in one spatial and one temporal dimensional space-time. The classical action of the lattice model in two-dimensional Minkowski space-time is  
\begin{align}
S &=  \frac{1}{2} \int_{-\infty}^\infty d t \int_{0}^L d x  \left [  \frac{\partial \phi^* }{\partial t} \frac{\partial \phi }{\partial t} -   \frac{\partial \phi^* }{\partial x} \frac{\partial \phi }{\partial x} -  m^2 | \phi |^2 \right ] \nonumber \\
& - \frac{1}{4 !}   \int_{-\infty}^\infty d t \int_{0}^L d x d y |  \phi (x, t)   |^2V(x-y) |  \phi (y, t)   |^2, \label{Minkowskiaction}
\end{align}
where complex  $\phi (x,t)$  field operator describes a charged scalar particle of mass $m$, and it satisfies the periodic boundary condition
\begin{equation}
\phi (x + L,t) = \phi (x,t) .
\end{equation}
The short-range spatially symmetric instantaneous interaction potential is represented by $V(x) = V(-x)$.

\subsection{Two-particle correlation function and its spectral representation}
The two charged scalar particles interaction can be studied  via evaluating time dependence of correlation function.  The    two-particle correlation function is defined by
\begin{align}
C (rt; r'0) & = \theta (t) \langle 0 |   \mathcal{O} (r, t )     \mathcal{O}^\dag (r' ,0)   | 0 \rangle \nonumber \\
& + \theta (-t) \langle 0 |    \mathcal{O}^\dag (r' ,0)    \mathcal{O} (r, t )   | 0 \rangle, 
\end{align}
 where  two identical charged particles creation operator after projecting out center of mass motion (CM) in the rest frame is given by
 \begin{equation}
    \mathcal{O}^\dag (r ,t) =  \frac{1}{\sqrt{2}} \int_0^L \frac{d x_2}{\sqrt{L}} \phi(r+x_2,t) \phi(x_2,t). \label{reltwoparticleoperator}
 \end{equation}
The factor $1/\sqrt{2}$ takes into account the exchange symmetry of two distinguishable charged particles. Inserting complete energy basis, $\sum_n | E_n \rangle \langle E_n | = 1$, between interpolating operators, and defining  two-particle relative wave function by 
\begin{equation}
\langle E_n  |   \mathcal{O}^\dag (r' ,0) | 0\rangle  = \frac{1}{\sqrt{L}} \frac{ \psi^{(L)*}_{E_n} (r')}{E_n}  ,   
\end{equation}
similarly   
\begin{equation}
\langle E_n  |   \mathcal{O}  (r ,t) | 0\rangle  = \frac{1}{\sqrt{L}} \frac{ \psi^{(L)*}_{E_n} (r)}{E_n} e^{ i  E_n t}
\end{equation}
defines the wave function of two antiparticles states, where we have assumed that   wave functions of two-particle and two-antiparticle are identical.
   The spectral representation of two-particle correlation function is thus given by
\begin{align}
C(rt; r'0) & = \frac{\theta (t)}{L} \sum_n \frac{ \psi^{(L)}_{E_n} (r)}{E_n} \frac{ \psi^{(L)*}_{E_n} (r')}{E_n} e^{- i  E_n t} \nonumber \\
& +  \frac{\theta (-t)}{L} \sum_n  \frac{ \psi^{(L)}_{E_n} (r')}{E_n}  \frac{ \psi^{(L) *}_{E_n} (r)}{E_n}e^{ i  E_n t}. \label{twocorrspectral}
\end{align}
The first   and the second terms in Eq.(\ref{twocorrspectral}) describes two-particle states propagating forward in time and  two-antiparticle states propagating backward in time respectively.
In general, both parities  contribute to energy states, for scalar particles considered in this work, only even parity energy states   survived due to Bose symmetry,
\begin{equation}
 \psi^{(L)}_{E_n} (-r) =  \psi^{(L)}_{E_n} (r).
\end{equation}
The  non-interacting    correlation function is     given by
\begin{equation}
C_0 (rt; r'0)  = \frac{1}{L} \sum_{p = \frac{2\pi n}{L}, n \in \mathbb{Z}} \frac{  \cos ( p r ) }{ E^{(0)}_p } \frac{  \cos (p r') }{E^{(0)}_p } e^{- i  E^{(0)}_p | t| },
\end{equation}
where free two-particle energies are $E^{(0)}_p  =2  \sqrt{p^2 + m^2}$  with $p = \frac{2\pi n}{L}, n \in \mathbb{Z}$.

The  two-particle relative wave function is required to satisfy relativistic Lippmann-Schwinger (LS) like equation, see e.g.  Appendix \ref{scattsolutionsinfvol} and \ref{RPAappendix},
\begin{equation}
 \psi^{(L)}_{E} (r) =  \int_0^L d r' G_0^{(L)} (r- r' ; E) V (r')   \psi^{(L)}_{E} (r'). \label{finitevolumeLSeq}
\end{equation}
 The relativistic finite volume Green's function is defined by, see e.g. Refs.~\cite{Guo:2019ogp,Guo:2020kph},
\begin{equation}
G_0^{(L)} (r; E) = \frac{1}{L} \sum_{q = \frac{2\pi n}{L}, n \in \mathbb{Z}} \frac{  1}{ \omega_q   } \frac{e^{i q r}}{E^2 - ( 2 \omega_q )^2}, \label{finitevolumefreeGreen}
\end{equation}
where $ \omega_q = \sqrt{q^2 + m^2}$ is the energy of a single particle with momentum  $q$. The relativistic wave function is normalized in momentum space by
\begin{equation}
\frac{1}{L} \sum_{p = \frac{2\pi n}{L}, n \in \mathbb{Z}} \frac{1}{2 \omega_p} \widetilde{\psi}^{(L)}_{E_n} (p)  \widetilde{\psi}^{(L)*}_{E_{n'}} (p) = E_n L \frac{ \delta_{n, n'}  +  \delta_{n, -n'}}{2}, \label{normalizationwavfunc}
\end{equation}
where the Fourier transform of wave function is defined by
\begin{equation}
 \frac{  \widetilde{\psi}^{(L)}_{E_n} (p) }{2 \omega_p}= \int_0^L d r  \psi^{(L)}_{E_n} (r) e^{i p r}.
\end{equation}

Using normalization relation of wave function in Eq.(\ref{normalizationwavfunc}), we find
\begin{equation}
  \frac{1}{L}  \sum_{p = \frac{2\pi n}{L}, n \in \mathbb{Z}} 2 \omega_p  \widetilde{C}(pt; p0) =  \sum_{n =0}^\infty \frac{ e^{- i  E_n | t | } }{E_n} , \label{integratedCt}
\end{equation}
where the Fourier transform of correlation function is defined by
\begin{equation}
 \widetilde{C}(pt; p'0) =  \int_0^L d r d r'  e^{i p r}   C(rt; r'0)  e^{- i p' r'}. \label{FourierTFdef}
\end{equation}
The Eq.(\ref{integratedCt}) may be considered as integrated correlation function in momentum space.
The difference of interaction and non-interacting integrated correlation functions is thus given  by
\begin{align}
\triangle C(t)   &= \frac{1}{L} \sum_{p = \frac{2\pi n}{L}, n \in \mathbb{Z}}   2 \omega_p  \left [   \widetilde{C}(p t; p 0) -   \widetilde{C}_0(p t; p 0)  \right ] \nonumber \\
& =  \sum_{n =0}^\infty \left [ \frac{ e^{- i  E_n t} }{E_n}  -  \frac{ e^{- i  E^{(0)}_n t} }{E^{(0)}_n}  \right ] .  \label{diffintegratedcorrfuncdef}
\end{align}

\subsection{Relating integrated correlation functions to scattering phase shift}

In momentum space, two-particle correlation function along diagonal direction is given by
\begin{align}
&  \widetilde{C}(p t; p0)  \nonumber \\
&= \frac{1}{L} \sum_n   \frac{1}{E_n^2  }  \frac{ \widetilde{\psi}^{(L)}_{E_n} (p)}{E_p} \frac{ \widetilde{\psi}^{(L)*}_{E_n} (p)}{E_p}    \left [\theta (t)  e^{- i  E_n t} + \theta(-t)  e^{ i  E_n t} \right ].
\end{align}
Using identity
\begin{equation}
i \int_{-\infty}^{\infty} \frac{d \lambda}{2\pi} \frac{e^{- i \lambda t}}{\lambda + i  0 } = \theta (t),  
\end{equation}
the particles time forward and antiparticles time backward propagations can be combined together.   The momentum space spectral representation of two-particle correlation function  along diagonal direction is thus expressed in terms of two-particle Green's function by
\begin{equation}
\widetilde{C}(p t; p0)  =    i \int_{-\infty}^{\infty} \frac{d \lambda}{2\pi}  \widetilde{G}^{(L)} (p ,  p; \lambda)  e^{- i \lambda t}    .
\end{equation}
The diagonal terms of momentum space two-particle Green's function is   defined by
\begin{equation}
\widetilde{G}^{(L)}(p, p; E)   =      \int_0^L d r  d r' e^{i p r} G^{(L)}(r, r'; E)  e^{- i p r'} ,
\end{equation}
where  two-particle Green's function is given by
\begin{equation}
G^{(L)}(r, r'; E)   =      \frac{1}{L} \sum_{q = \frac{2\pi n}{L}, n\in \mathbb{Z} } \frac{1   }{\omega_q}   \frac{    \psi^{(L)}_{E_q} (r)   \psi^{(L)*}_{E_q} (r') }{  E^2 -  E_q^2  + i 0 }   ,
\end{equation}
and $E_q = 2 \omega_q = 2 \sqrt{q^2+m^2}$.
Hence the difference of integrated correlation functions is given in terms of Green's functions by
\begin{align}
 \triangle C(t)  &= i \int_{-\infty}^{\infty} \frac{d \lambda}{2\pi}   e^{- i \lambda t}   \nonumber \\
&  \times   \frac{1}{L} \sum_{p = \frac{2\pi n}{L}, n \in \mathbb{Z}}   2 \omega_p \left [\widetilde{G}^{(L)}(p, p; \lambda)-\widetilde{G}^{(L)}_0(p, p; \lambda)  \right ] ,
\end{align}
where the Fourier transform of Green's functions are defined in a similar way as in Eq.(\ref{FourierTFdef}).

Next, using relativistic Friedel formula relation in infinite volume, see Appendix \ref{relatFriedelformula},
    \begin{align} 
  &  \int_{-\infty}^\infty   \frac{d p}{2\pi}     \omega_p   \left [  \widetilde{G}^{(\infty)} (p , p; E)    -  \widetilde{G}^{(\infty)}_0 (p , p; E)  \right ]  \nonumber \\
  &  = -       \frac{1}{\pi} \int_{4 m^2}^\infty d s \frac{ \delta(\sqrt{s}) }{  (s - E^2 - i 0 )^2}    ,
\end{align}
where $\delta (E)$ is the scattering phase shift of two scalar particles, at large volume limit, the difference of integrated correlation functions thus approaches
\begin{align}
 \triangle C(t)  & \stackrel{L \rightarrow \infty}{\rightarrow}     -    \frac{1}{\pi} \int_{4 m^2}^\infty d s \delta(\sqrt{s})  \left [    i \int_{-\infty}^{\infty} \frac{d \lambda}{\pi}     \frac{  e^{- i \lambda t} }{ (s - \lambda^2 - i 0)^2}  \right ]  .
\end{align}
Completing integration in bracket, a compact form of large volume limit of difference of integrated correlation functions can be found,
\begin{equation}
 \triangle C(t)   \stackrel[t=-i \tau]{L \rightarrow \infty}{\rightarrow}    -     \frac{1}{\pi} \int_{2 m}^\infty d \epsilon   \delta( \epsilon)   \frac{d}{d \epsilon}  \left (  \frac{e^{ -   \epsilon \tau} }{ \epsilon}   \right ) .
\end{equation}
This relation  that is also listed in Eq.~(\ref{mainresult}) is our main result. Using an exactly solvable model and perturbation theory in Sec.~\ref{exactmodelcontactpot} and Sec.~\ref{perturbationresult} respectively, we show that  the difference of integrated correlation functions converges rapidly to its infinite volume limit.

\subsection{Exactly solvable model with a contact interaction}\label{exactmodelcontactpot}
Considering a contact interaction, 
\begin{equation}
V(r) = V_0 \delta(r),
\end{equation}
the action in Eq.(\ref{Minkowskiaction}) is thus reduced to a complex scalar $\phi^4$ theory action. The finite volume LS equation in Eq.(\ref{finitevolumeLSeq}) yields  the quantization condition
\begin{equation}
 \frac{1}{V_0}  =   G_0^{(L)} ( 0 ; E)  ,
\end{equation}
where finite volume free two-particle Green's function is defined in Eq.(\ref{finitevolumefreeGreen}).

\begin{figure*}
 \centering
 \begin{subfigure}[b]{0.47\textwidth}
\includegraphics[width=0.99\textwidth]{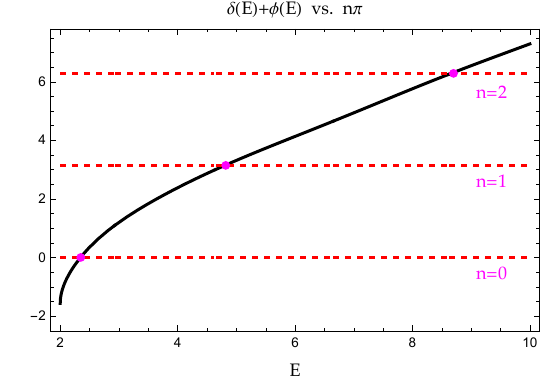}
\caption{  }\label{qcplot}
\end{subfigure} 
\begin{subfigure}[b]{0.49\textwidth}
\includegraphics[width=0.99\textwidth]{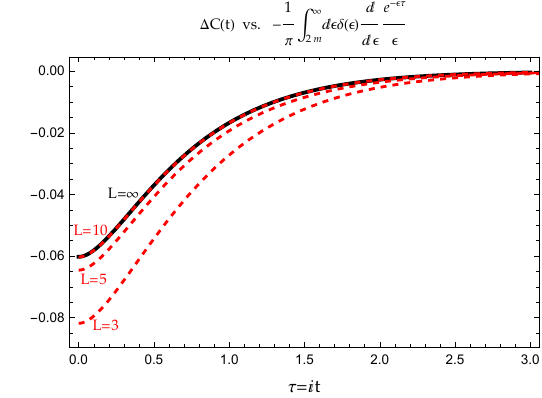}
\caption{    }\label{modeldctplot}
\end{subfigure}
 \caption{The energy spectra and difference of integrated correlation function plots: (a) $\delta(\epsilon_n) +\phi(\epsilon_n)$ (solid black) vs. $n \pi$ (dashed red) with $L =3$, energy spectra are located at intersection points of black and red curves; (b) $  - \frac{1}{\pi}    \int_{2m}^\infty   d  \epsilon  \delta (\epsilon)  \frac{d}{d \epsilon} \left (   \frac{   e^{-    \epsilon  \tau}  }{\epsilon} \right )  $  (solid black) vs. $\triangle C (t)  =  \sum_{n=0}^\infty \left [ \frac{e^{- E_n \tau}}{E_n} - \frac{e^{- E^{(0)}_n \tau}}{E^{(0)}_n}  \right ]  $ (dashed red)  with $L=3,5,10$. The rest of parameters are taken as:       $m=1$ and $V_0=5$.  \label{boxfigs} }
 \end{figure*}

Using scattering solutions in infinite volume, Eq.(\ref{scattamp}) and Eq.(\ref{scattampparametrization}) in Appendix \ref{scattsolutionsinfvol}, the potential strength $V_0$ is related to infinite volume free particles Green's function by
\begin{equation}
 \frac{1}{V_0}   = Re \left [ G_0^{(\infty)} ( 0 ; E)  \right ] - \rho(E) \cos \delta(E), \label{phaseshift}
\end{equation}
where  the analytic expression of $G_0^{(\infty)} ( 0 ; E)$ and $\rho(E)$ are given  in Eq.(\ref{inffreeGreenanaly}) and Eq.(\ref{infphasefactor}) respectively, see details in Appendix \ref{scattsolutionsinfvol}. The quantization condition can be rewritten in a form that is known as L\"uscher  formula,
\begin{equation}
  \cos \delta(E) =  \frac{Re \left [ G_0^{(\infty)} ( 0 ; E)  \right ]  - G_0^{(L)} ( 0 ; E) }{\rho(E)}, \label{lusherformula}
\end{equation}
the right hand side of Eq.(\ref{lusherformula}) is typically referred to zeta function that is associated to the long range geometry of a lattice, see e.g. Ref.\cite{Luscher:1990ux}. Let's rearrange L\"uscher  formula to
\begin{equation}
 \delta(E_n)  + \phi (E_n) = n \pi.  \label{qccontactpot}
\end{equation}
where subscript-$n$ in $E_n$ is used to label the $n$-th eigen-energy of the system, and
\begin{equation}
 \phi (E) = -\cot^{-1} \left [ \frac{Re \left [ G_0^{(\infty)} ( 0 ; E)  \right ]  - G_0^{(L)} ( 0 ; E) }{\rho(E)} \right ]   + l \pi, 
\end{equation}
where $l \in \mathbb{Z}$. 
The $ l \pi $ is added to keep $\phi(E)$ as a monotonic function  and prevent  jumping  at branch points when $E=2 \sqrt{(\frac{2\pi n}{L})^2 + m^2 }$ where $n\in \mathbb{Z}$, see e.g. Fig.~\ref{qcplot}.

The fast convergence of   relation 
   \begin{equation}
    \sum_{n=0}^\infty \left [ \frac{e^{- E_n \tau}}{E_n} - \frac{e^{- E^{(0)}_n \tau}}{E^{(0)}_n}  \right ]     \stackrel{L \rightarrow \infty}{\rightarrow}  - \frac{1}{\pi}    \int_{2m}^\infty   d  \epsilon  \delta (\epsilon)  \frac{d}{d \epsilon} \left (   \frac{   e^{-    \epsilon  \tau}  }{\epsilon} \right )    
\end{equation}
can be verified   numerically, where interacting energy levels are determined by quantization condition in Eq.(\ref{qccontactpot}). The free particles energy levels are given by $E_n^{(0)}  =2 \sqrt{ (\frac{2\pi n}{L})^2 + m^2 }$. The phase shift is computed by using Eq.(\ref{phaseshift}). The difference of finite volume integrate correlation functions approaches its infinite volume limit rapidly, see e.g. Fig.~\ref{modeldctplot}.

\subsection{Leading order result of perturbation calculation}\label{perturbationresult}

The fast convergence of the difference of integrated correlation functions to its infinite volume limit can also be checked straightforwardly by perturbation theory. The perturbation calculation can be carried out in a similar way as demonstrated in Ref.~\cite{Guo:2023ecc},  the leading order contribution is given by
\begin{align}
& C ( r, t; r', 0 )  -  C_0 ( r, t; r' ,0 )  \nonumber \\
&  =  - \frac{ i V_0}{L} \int_{0}^L  d x_2  \int_{0}^L  d x'_2  \int_{-\infty}^\infty d t' \int_0^L d x''   \nonumber \\
& \times    D_0^{-1} (r+ x_2 -x'' , t - t''  ) D^{-1}_0 (x_2 -x'' , t -t'' )   \nonumber \\
& \times    D_0^{-1} (x'' - r'-  x'_2 ,  t''  ) D^{-1}_0 (x'' -x'_2 ,   t'' )+ \mathcal{O} (V_0^2) , \label{perturbationcorrelation}
\end{align}
where free two-particle propagator   is defined by
\begin{equation}
D_0^{-1} (x   , t )   = i \int_{-\infty}^\infty  \frac{d  \epsilon }{2\pi} \frac{1}{L} \sum_{k = \frac{2\pi n}{L}, n \in \mathbb{Z}}  \frac{  e^{i k x  }  e^{i  \epsilon   t   }}{ \epsilon^2  -( k^2 +  m^2 )} . \label{invD0expression}
\end{equation}
Carrying out space and time integration, we find
\begin{align}
& C ( r, t; r', 0 )  -  C_0 ( r, t; r' ,0 )  \nonumber \\
&  = -  i V_0   \int_{-\infty}^\infty  \frac{d \epsilon }{2\pi}  e^{i \epsilon   t   } G_0^{(L)} (r; \epsilon ) G_0^{(L)} (r'; \epsilon) + \mathcal{O} (V_0^2)  ,
\end{align}
where the finite volume free particles Green's function, $G_0^{(L)} (r; \epsilon ) $, is defined in Eq.(\ref{finitevolumefreeGreen}). Using definition of the difference of integrated correlation function in Eq.(\ref{diffintegratedcorrfuncdef}) and further carrying out the integration of $  \epsilon $,   the leading order result of perturbation calculation  is given by
\begin{equation}
 \triangle C (  t )    \stackrel{t=- i \tau}{ =}  -      \frac{V_0}{L}  \sum_{k = \frac{2\pi n}{L}, n \in \mathbb{Z}}     \frac{  \tau +  \frac{1}{E_k }   }{  E_k^3 }     e^{- E_k \tau}    + \mathcal{O} (V_0^2)    ,  \label{perturbfinitevolume}
\end{equation}
where again $E_k =2 \sqrt{k^2+ m^2}$ is total energy of two particles. At the large volume limit, it thus approaches
 \begin{equation}
 \triangle C (  t )   \stackrel[t=-i\tau]{L \rightarrow \infty}{\rightarrow}  -      V_0  \int_{- \infty}^{\infty} \frac{d k}{2\pi}    \frac{  \tau +  \frac{1}{E_k }   }{  E_k^3 }     e^{- E_k \tau}    + \mathcal{O} (V_0^2)    .
\end{equation}

On the other hand,  using the Taylor expansion of scattering phase shift,
\begin{equation}
\delta(E_k ) = -  \frac{V_0}{4 k E_k }  +  \mathcal{O} (V_0^2) ,
\end{equation}
we thus find
 \begin{align}
&  -   \frac{1}{\pi}   \int_{2m}^\infty   d  \epsilon  \delta (\epsilon)  \frac{d}{ d \epsilon}  \left (   \frac{   e^{-    \epsilon  \tau}  }{\epsilon}  \right ) \nonumber \\
& = -   V_0     \int_{-\infty}^\infty  \frac{  d k}{2\pi    }    \left ( \tau+ \frac{1}{\epsilon_k } \right )   \frac{   e^{-    \epsilon_k  \tau}  }{\epsilon^3_k}  +  \mathcal{O} (V_0^2) ,
\end{align}
 where $\epsilon_k = 2 \sqrt{k^2+ m^2}$.  This is indeed consistent with perturbation calculation.

\section{Monte Carlo Simulation of complex \texorpdfstring{$\phi^4$}{phi**4} lattice model}\label{MCsimulation}

In this section, the formalism is tested by carrying out Monte Carlos simulation of complex $\phi^4$ lattice model, which describes charged scalar particles interacting with a contact potential. The results are compared with the result by using  L\"uscher formula.  The Euclidean spacetime lattice $\phi^4$ action is  given by, see e.g. Ref.~\cite{Guo:2018xbv}, 
\begin{align} 
S_E  =& -   \kappa  \sum_{x, \tau, \hat{n}_x, \hat{n}_\tau  } \hat{\phi}^*(x, \tau) \hat{\phi}(x+\hat{n}_x, \tau+\hat{n}_\tau )  + c.c. \nonumber \\
&+ (1-2 \lambda)  \sum_{x, \tau} | \hat{\phi}(x,\tau) |^2 + \lambda \sum_{x,\tau} |\hat{ \phi}(x,\tau) |^4 , \label{Euclideanaction}
\end{align}
where  $(x, \tau)$    refer to discrete coordinates of   Euclidean $L \times T$  lattice site:   $    x \in [0, L-1]$ and $\tau \in [0,T-1]$. The lattice spacing, $a$, is set to unity.   The $(\hat{n}_x, \hat{n}_\tau )$ denotes the unit vector in direction $(x,\tau)$ on a  periodic square  lattice.  The parameters $ (\kappa, \lambda)$ are related to bare mass $m_0$ and bare coupling constant  $g_0$ of interacting term $\frac{g_0}{4!} | \phi |^4$ by   $m_0^2 = \frac{1-2 \lambda}{\kappa}-8$ and $g_0=\frac{6 \lambda}{\kappa^2}$, see Ref.~\cite{Guo:2018xbv}.   The $\phi$ field in  $\phi^4$  lattice model has been rescaled:  $$\phi (x,\tau)=  \sqrt{2\kappa } \hat{\phi} (x,\tau).  $$  The numerical simulation is carried out by Hybrid Monte Carlo (HMC) algorithm, the   details of HMC algorithm is described in Ref.~\cite{Guo:2018xbv}.

 The single particle and two-particle  correlation functions are defined respectively by
\begin{equation}
 C^{(\phi)}(x \tau, x' 0) =   \frac{   \int \mathcal{D} \phi \mathcal{D} \phi^\dag   \phi (x, \tau)  \phi^\dag ( x',  0)    e^{-S_E }  }{    \int \mathcal{D} \phi \mathcal{D} \phi^\dag     e^{-S_E }},
\end{equation}
and  
\begin{equation}
C^{(2\phi)} (r \tau, r' 0) =   \frac{   \int \mathcal{D} \phi \mathcal{D} \phi^\dag   \mathcal{O} (r, \tau)  \mathcal{O}^\dag ( r',  0)    e^{-S_E }  }{    \int \mathcal{D} \phi \mathcal{D} \phi^\dag     e^{-S_E }},
\end{equation}
where the relative motion of two-particle interpolating operator  is defined in Eq.(\ref{reltwoparticleoperator}).

The individual energy levels can be projected out by
\begin{equation}
 \widetilde{C}^{(\phi,2\phi)}( p, \tau) =  \sum_{ x, x'  \in [0, L-1]} e^{i p x} C^{(\phi,2\phi)}(x \tau, x' 0) e^{- i p x'} ,
\end{equation}
where $p = \frac{2\pi n}{L}, n \in  [ - \frac{L}{2} +1, \frac{L}{2}]$.

  \begin{figure*}
 \centering
 \begin{subfigure}[b]{0.45\textwidth}
\includegraphics[width=0.99\textwidth]{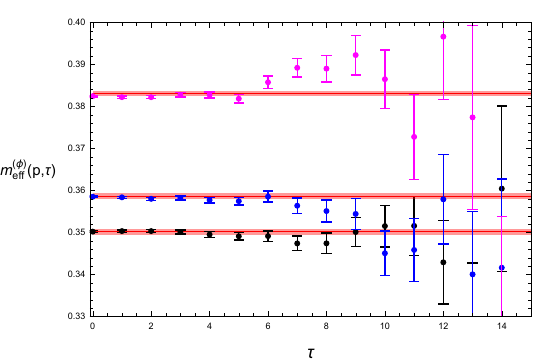}
\caption{ single particle effective mass.}\label{onefreemassplot}
\end{subfigure} 
\begin{subfigure}[b]{0.46\textwidth}
\includegraphics[width=0.99\textwidth]{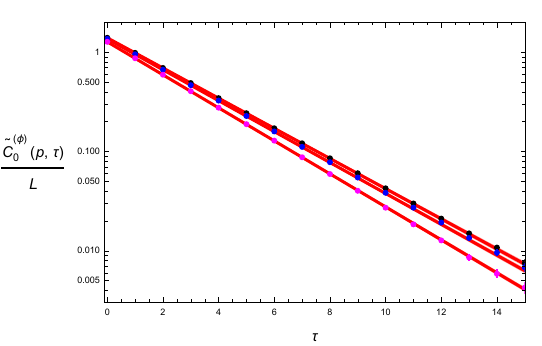}
\caption{  $ \widetilde{C}^{(\phi)}_0( p, \tau)/L $ vs. lattice data.}\label{onefreeCtqsplot}
\end{subfigure}
 \begin{subfigure}[b]{0.45\textwidth}
\includegraphics[width=0.99\textwidth]{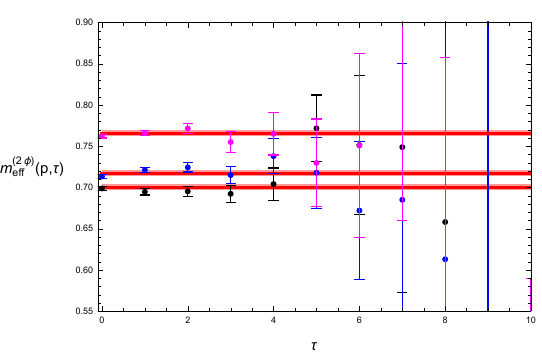}
\caption{ two-particle effective mass.}\label{twofreemassplot}
\end{subfigure} 
\begin{subfigure}[b]{0.46\textwidth}
\includegraphics[width=0.99\textwidth]{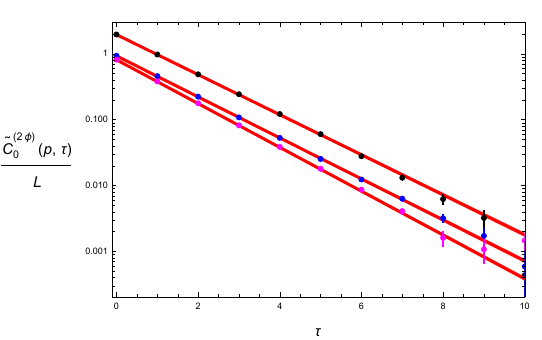}
\caption{  $ \widetilde{C}^{(2\phi)}_0( p, \tau)/L $ vs. lattice data.}\label{twofreeCtqsplot}
\end{subfigure}
\caption{ Comparison of   (a,c)  single particle and two particles effective mass  $m^{(\phi,2\phi)}_{eff} (p,\tau) = \ln \frac{ \widetilde{C}^{(\phi,2\phi)}_0( p, \tau) }{ \widetilde{C}^{(\phi,2\phi)}_0( p, \tau+1)   }$ (red band) vs.  lattice data,  where $p = \frac{2\pi n}{L}$, $n =0$ (black error bars), 1 (blue error bars), and 2 (purple error bars);    (b,d) $ \frac{\widetilde{C}^{(\phi,2\phi)}_0( p, \tau)}{L} $ (red band)  vs.  lattice data, where $p = \frac{2\pi n}{L}$, $n =0$ (black error bars), 1 (blue error bars), and 2 (purple error bars).
The model parameters are taken as:    $\kappa =0.1213$,   $\lambda  = 0$,    $T=100$ and $ L=80$. }\label{freeoneCtplot1}
\end{figure*}

  \begin{figure*}
 \centering
 \begin{subfigure}[b]{0.45\textwidth}
\includegraphics[width=0.99\textwidth]{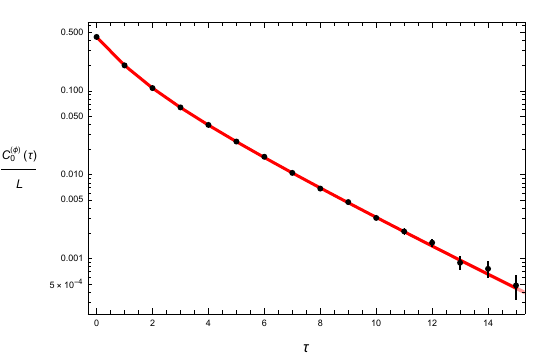}
\caption{  $ \frac{C^{(\phi)}_0(  \tau)}{L} $ vs. lattice data.}\label{onefreeL80Ctplot}
\end{subfigure} 
\begin{subfigure}[b]{0.45\textwidth}
\includegraphics[width=0.99\textwidth]{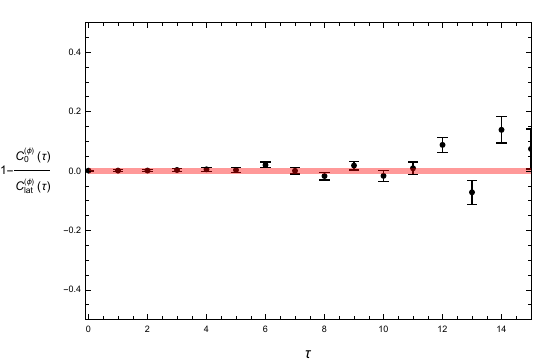}
\caption{  $ 1-  C^{(\phi)}_0(  \tau)/C^{(\phi)}_{lat}(  \tau)$.}\label{onefreeL80diffCtplot}
\end{subfigure}
 \begin{subfigure}[b]{0.44\textwidth}
\includegraphics[width=0.99\textwidth]{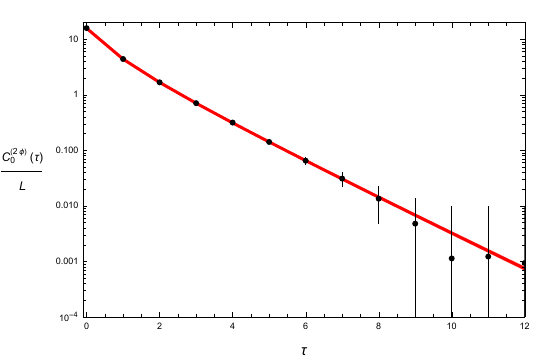}
\caption{  $ \frac{C^{(2\phi)}_0(  \tau)}{L} $ vs. lattice data.}\label{twofreeL80Ctqplot}
\end{subfigure} 
\begin{subfigure}[b]{0.46\textwidth}
\includegraphics[width=0.99\textwidth]{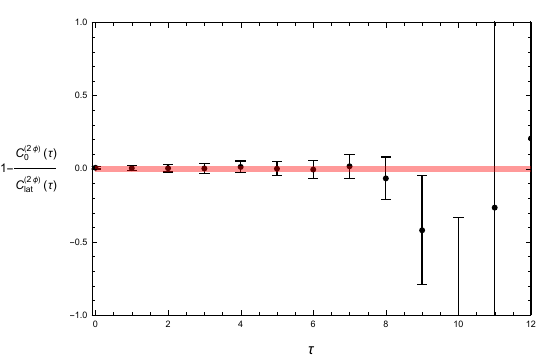}
\caption{  $ 1-  C^{(2\phi)}_0(  \tau)/C^{(2\phi)}_{lat}(  \tau)$.}\label{twofreeL80diffCtqplot}
\end{subfigure}
\caption{ (a,c) Comparison of    $\frac{C^{(\phi,2\phi)}_0(  \tau)}{L}$ (red band)  vs. lattice data (black error bars);     (b,d) Plot of $ 1-  C^{(\phi,2\phi)}_0(  \tau)/C^{(\phi,2\phi)}_{lat}(  \tau)$ (black error bars), red  error band is also plotted. 
The model parameters are taken as:       $\kappa =0.1213$,   $\lambda  = 0$,   $T=100$ and $ L=80$. }\label{freeoneCtplot2}
\end{figure*}

\subsection{Simulation test for non-interacting case: \texorpdfstring{$\lambda =0$}{lambda=0.8}}

For non-interacting particles by setting $\lambda =0$ in Euclidean action in Eq.(\ref{Euclideanaction}),  analytic expression of correlation functions can be found.   The single particle correlation is given by
\begin{equation}
    C_0^{(\phi)}(x \tau, x' 0)     =   \frac{1}{L} \sum_{k = \frac{2\pi n}{L} }^{n  \in   [- \frac{L}{2} +1, \frac{L}{2}] }    e^{i k (x-x') }  G_{\phi}  (k, \tau)   ,
\end{equation}
and two particles correlation function is given by
\begin{align}
 &   C_0^{(2\phi)}(r \tau, r' 0)   \nonumber \\
 &  =\frac{1}{L} \sum_{k = \frac{2\pi n}{L} }^{n \in  [- \frac{L}{2} +1, \frac{L}{2}]  }  \cos (k r) \cos (k r')     \left [  G_{\phi}  (k, \tau)   \right ]^2  ,
\end{align}
where
\begin{equation}
G_{\phi }  (k, \tau) = \frac{1}{T} \sum_{\omega = \frac{2\pi n}{T}}^{n \in [0,T-1]}      \frac{ e^{i \omega \tau   }   }{ 2 -2  \cos \omega     - 2   \cos   k   + 2 \cosh m    }  .
\end{equation}
In above expressions,  the lattice spacing $a$ has been set to unity, the lattice spacing can be installed easily by, such as replacing $  \omega$ by dimensionless argument in $\cos (a \omega)$.
 At the  limit of $T \rightarrow \infty$ and also taking lattice spacing  to zero ($a\rightarrow 0$),   we find
\begin{equation}
G_{\phi }  (k, \tau)  \stackrel[a\rightarrow 0]{T \rightarrow \infty}{ \rightarrow }   \frac{e^{- \omega_k \tau}  }{2 \omega_k }   ,
\end{equation}
 and  single and two-particle correlation functions approach
\begin{equation}
   C_0^{(\phi)}(x \tau, x' 0)  \stackrel[a\rightarrow 0]{T \rightarrow \infty}{ \rightarrow }  \frac{1}{L} \sum_{k = \frac{2\pi n}{L} }^{n \in \mathbb{Z}}  e^{i k (x-x') }    \frac{e^{- \omega_k \tau}  }{2 \omega_k } ,  
\end{equation}
and
\begin{equation}
   C_0^{(2 \phi)}(x \tau, x' 0)  \stackrel[a\rightarrow 0]{T \rightarrow \infty}{ \rightarrow }  \frac{1}{L} \sum_{k = \frac{2\pi n}{L} }^{n \in \mathbb{Z}}  \cos (k r) \cos (k r')  \frac{e^{-2 \omega_k \tau} }{ (2 \omega_k)^2 }    ,  
\end{equation}
where $\omega_k= \sqrt{k^2+m^2}$.

The individual energy level can be extracted by projecting the correlation functions into momentum space along diagonal direction, 
\begin{equation}
 \widetilde{C}_0^{(\phi,2\phi)}( p, \tau) =  \sum_{ x, x'  \in [0, L-1]} e^{i p x} C_0^{(\phi,2\phi)}(x \tau, x' 0) e^{- i p x'} ,
\end{equation}
thus we find
\begin{equation}
   \frac{1}{L} \widetilde{C}^{(\phi)}_0( p, \tau)   =  G_{ \phi }  (p, \tau)    \stackrel[a\rightarrow 0]{T \rightarrow \infty}{ \rightarrow }   \frac{e^{- \omega_p \tau}  }{2 \omega_p }  , \label{projCphipt}
\end{equation}
 where $p =\frac{2\pi n}{L}, n \in [ -\frac{L}{2} +1 , \frac{L}{2}]$,  and
\begin{equation}
  \frac{1}{L}  \widetilde{C}_0^{(2\phi)}(p, \tau)  =  \frac{\sigma_p}{2}   \left [  G_{\phi}  (p, \tau)   \right ]^2    \stackrel[a\rightarrow 0]{T \rightarrow \infty}{ \rightarrow }    \frac{\sigma_p}{2}    \frac{e^{- 2 \omega_p \tau}  }{ (2 \omega_p)^2 } , \label{projCtwophipt}
\end{equation}
where the symmetry factor $\sigma_p$ is defined by
\begin{equation}
\sigma_p = \begin{cases} 2 , & if \ \ p=0, \\ \frac{1}{2}, & if \ \ p = \pi,  \\ 1 , &  otherwise.   \end{cases}
\end{equation}

At the limit of $a \rightarrow 0$ and $T \rightarrow \infty$, the  integrated correlation functions  of free particles approach
\begin{equation}
C^{(\phi)}_0(  \tau) =   \frac{1}{L} \sum_{p = \frac{2\pi n}{L}}^{n \in   [- \frac{L}{2} +1, \frac{L}{2}] } \widetilde{C}^{(\phi)}_0( p, \tau)    \stackrel[a\rightarrow 0]{T \rightarrow \infty}{ \rightarrow }     \sum_{ p = \frac{2\pi n}{L}}^{n \in \mathbb{Z}}    \frac{e^{- \omega_p \tau}  }{2 \omega_p } , \label{integratedCphit}
\end{equation}
and
\begin{align}
   C^{(2 \phi)}_0(  \tau)  & =   \frac{1}{L} \sum_{p = \frac{2\pi n}{L}}^{n \in   [ -\frac{L}{2} +1 , \frac{L}{2}] }  2 \omega^{(a, L)}_p  \widetilde{C}_0^{(2\phi)}(p, \tau)      \nonumber \\
   & \stackrel[a\rightarrow 0]{T \rightarrow \infty}{ \rightarrow }    \sum_{ p = \frac{2\pi n}{L}}^{n \in  [0, \infty] }  \frac{e^{- 2 \omega_p \tau}  }{ 2 \omega_p} , \label{integratedCtwophit}
\end{align}
where 
\begin{equation}
\omega^{(a, L)}_p =\cosh^{-1} \left [ 1 + \cosh m - \cos p \right ]  \stackrel{a\rightarrow 0}{ \rightarrow }   \sqrt{p^2+ m^2}.  \label{omegafiniteaandL}
\end{equation}

The simulations for non-interacting charged scalar particles are performed with the choice of the parameters:  $\kappa =0.1213$, and $\lambda  = 0$.  The temporal extent of the lattice  and the   spatial extent of lattice is fixed at $T=100$ and $L=80$ respectively. For each set of lattice,   one million measurements are generated.

The mass of single  particle is measured by fitting projected single particle correlation function $ \widetilde{C}^{(\phi)}_0( p, \tau) $ with $p=0$, we find $m = 0.3502\pm 0.0032$. Using the single particle's mass  as input, the comparison  of analytic expression of effective mass,  where
\begin{equation}
m^{(\phi,2\phi)}_{eff} (p,\tau) = \ln \frac{ \widetilde{C}^{(\phi,2\phi)}_0( p, \tau) }{ \widetilde{C}^{(\phi,2\phi)}_0( p, \tau+1)   },
\end{equation}  
vs. lattice data  are plotted in Fig.~\ref{onefreemassplot} and Fig.~\ref{twofreemassplot}. The plots of  projected single particle and two-particle correlation functions $ \widetilde{C}^{(\phi,2\phi)}_0( p, \tau) $ that are defined in Eq.(\ref{projCphipt}) and Eq.(\ref{projCtwophipt})  vs. lattice data are shown in Fig.~\ref{onefreeCtqsplot} and  Fig.~\ref{twofreeCtqsplot}.   The  plots of integrated correlation functions, $  C^{(\phi,2\phi)}_0(  \tau)/L    $  that are defined in Eq.(\ref{integratedCphit}) and Eq.(\ref{integratedCtwophit}),   vs. lattice data are shown in Fig.~\ref{onefreeL80Ctplot} and Fig.~\ref{twofreeL80Ctqplot}.  We also plot  $ 1-  C^{(\phi,2\phi)}_0(  \tau)/C^{(\phi,2\phi)}_{lat}(  \tau)$  in Fig.~\ref{onefreeL80diffCtplot} and Fig.~\ref{twofreeL80diffCtqplot} to illustrate good agreement between lattice data and analytic expression of free single  particle and two particles correlation functions.

  \begin{figure*}
 \centering
 \begin{subfigure}[b]{0.45\textwidth}
\includegraphics[width=0.99\textwidth]{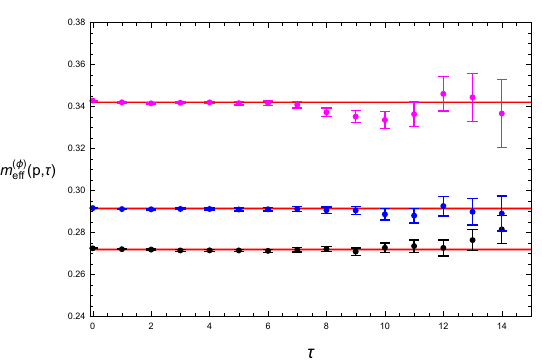}
\caption{ single particle effective mass  for $L=60$.}\label{onemassplot}
\end{subfigure} 
\begin{subfigure}[b]{0.45\textwidth}
\includegraphics[width=0.99\textwidth]{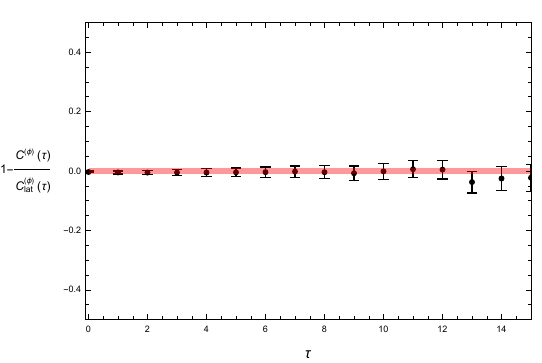}
\caption{  $ 1-  C^{(\phi)}(  \tau)/C^{(\phi)}_{lat}(  \tau)$  for $L=60$.}\label{oneL60diffCtplot}
\end{subfigure}
 \begin{subfigure}[b]{0.44\textwidth}
\includegraphics[width=0.99\textwidth]{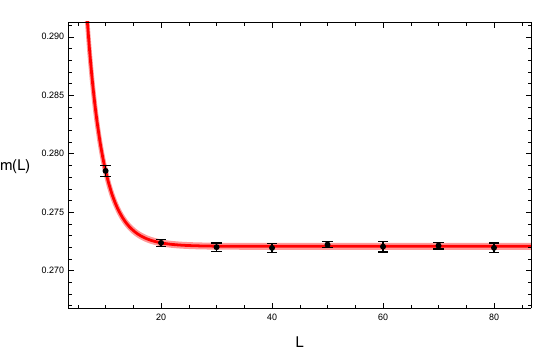}
\caption{ single particle mass  $m (L) =$ as a function of $L$.}\label{onemassallLplot}
\end{subfigure} 
\begin{subfigure}[b]{0.43\textwidth}
\includegraphics[width=0.99\textwidth]{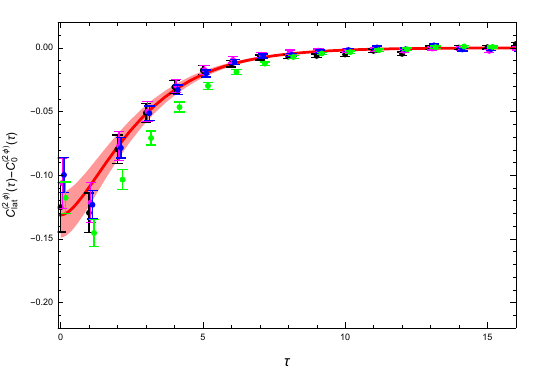}
\caption{  $  \triangle C^{(2\phi)}(\tau) $ vs. its infinite volume limit.}\label{twodiffCtqplot}
\end{subfigure}
\caption{   (a) Plot of single particle and two particles effective mass  $m^{(\phi)}_{eff} (p,\tau) = \ln \frac{ \widetilde{C}_{lat}^{(\phi)}( p, \tau) }{ \widetilde{C}_{lat}^{(\phi )}( p, \tau+1)   }$,  where $p = \frac{2\pi n}{L}$, $n =0$ (black error bars), 1 (blue error bars), and 2 (purple error bars). The red dashed lines represent free particles energy level, solid red lines and red bands are center value and its error band of interacting energy levels;    (b)  Plot of $ 1-  C^{(\phi)}_0(  \tau)/C^{(\phi)}_{lat}(  \tau)$ (black error bars), and red  error band is also plotted;  (c) Single particle mass as a function of lattice size $L$, $m(L) =m + \frac{c}{\sqrt{L}} e^{- m L}$, where $m = 0.272\pm 0.0015$ and $c = 0.31 \pm 0.05$; (d) Plot of $  \triangle C^{(2\phi)}(\tau) $  for  lattice data with various $L$'s ranging from  $L=10$(green), $20$(blue), $30$(magenta),    and $40$(black error bars)   vs. its infinite volume limit (solid red and red band). The lattice data   are plotted offsite horizontally  for better visualization.
The model parameters are taken as:    $\kappa =0.1286$, and $\lambda  = 0.01$ and    $T=120$. }\label{twoCtplot}
\end{figure*}

\subsection{Contact interacting cases: \texorpdfstring{$\lambda \neq 0$}{nonzero lambda}}

\subsubsection{Extracting phase shift from integrated correlation function}\label{V0bydCt}

The simulations for interacting charged scalar particles are performed with the choice of the parameters:  $\kappa =0.1286$, and $\lambda  = 0.01$.  The temporal extent of the lattice  is fixed at $T=120$,  and various spatial extents of lattice   $L$'s are computed.

The mass of single  particle is measured by fitting projected single particle correlation function $ \widetilde{C}^{(\phi)}_0( p, \tau) $ with $p=0$, we find 
$ m  \simeq  0.2720\pm 0.0015$. The example  of single particle   effective mass of lattice data are plotted in Fig.~\ref{onemassplot},  and single particle mass as a function of lattice size  $L$,  
\begin{equation}
m(L) =m + \frac{c}{\sqrt{L}} e^{- m L} ,
\end{equation}
 where $m = 0.272\pm 0.0015$ and $c = 0.31 \pm 0.05$,  in Fig.~\ref{onemassallLplot}.   The  plot of integrated single particle correlation function vs. data,   $ 1-  C^{(\phi )}_0(  \tau)/C^{(\phi )}_{lat}(  \tau)$,    is shown in Fig.~\ref{oneL60diffCtplot}. The difference of integrated two particles correlation functions between lattice data and non-interacting analytic expression, $  \triangle C^{(2\phi)}(\tau) $,  is plotted in Fig.~\ref{twodiffCtqplot}, where the  $  \triangle C^{(2\phi)}(\tau) $  is defined by
\begin{equation}
\triangle C^{(2 \phi)}(  \tau) = C^{(2 \phi)}_{lat}(  \tau)  - C^{(2 \phi)}_0(  \tau) ,
\end{equation}
and 
\begin{align}
   C^{(2 \phi)}_{lat}(  \tau)  & =   \frac{1}{L} \sum_{p = \frac{2\pi n}{L}}^{n \in  [- \frac{L}{2} +1, \frac{L}{2}] }  2 \omega^{(a, L)}_p  \widetilde{C}_{lat}^{(2\phi)}(p, \tau)    .
\end{align}
  The analytical expression of non-interacting  particles correlation function $C^{(2 \phi)}_0(  \tau) $ is defined in Eq.(\ref{integratedCtwophit}), and $\omega^{(a, L)}_p $ is defined in Eq.(\ref{omegafiniteaandL}).  The contact interaction coupling strength can be extracted from $  \triangle C^{(2\phi)}(\tau) $, see Fig.~\ref{twodiffCtqplot}, we find
  \begin{equation}
  V_0 =  0.196 \pm  0.03 0. 
  \end{equation}
Similar to non-relativistic case \cite{Guo:2023ecc},  both interacting and non-interacting particles correlation functions are divergent as $L \rightarrow \infty$ and $\tau \sim 0$, the divergent part behaves as $C^{(2 \phi)}_{lat}(  \tau \sim 0)  \propto  L \ln L$. The divergent parts are cancelled out so that $\triangle C^{(2 \phi)}(  \tau) $ remains well behaved as $L \rightarrow \infty$. The cancellation of divergence is crucial, hence the accurate representation of $ C^{(2 \phi)}_0(  \tau) $ is important. The consequence is that  $\triangle C^{(2 \phi)}(  \tau) $ is sensitive to the mass of $\phi$ field near small Euclidean time that ultimately generate large uncertainties near $\tau \sim 0$.

  \begin{figure*}
 \centering
 \begin{subfigure}[b]{0.45\textwidth}
\includegraphics[width=0.99\textwidth]{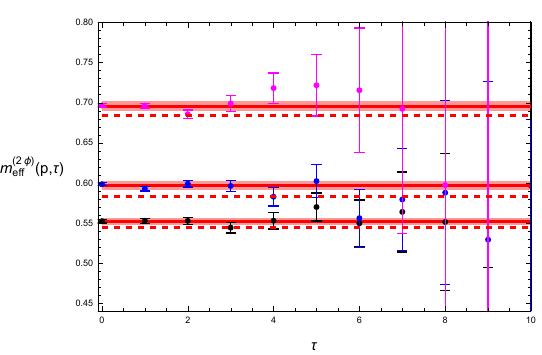}
\caption{ two-particle effective mass for $L=60$.}\label{twomassplot}
\end{subfigure} 
\begin{subfigure}[b]{0.43\textwidth}
\includegraphics[width=0.99\textwidth]{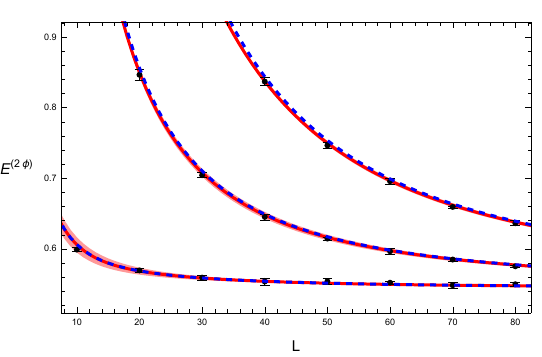}
\caption{  $  E^{(2\phi)}  $ for various $L$'s.}\label{luscherplot}
\end{subfigure}
\caption{   (a) Plot of   two particles effective mass  $m^{(2\phi)}_{eff} (p,\tau) = \ln \frac{ \widetilde{C}_{lat}^{(2\phi)}( p, \tau) }{ \widetilde{C}_{lat}^{( 2\phi)}( p, \tau+1)   }$,  where $p = \frac{2\pi n}{L}$, $n =0$ (black error bars), 1 (blue error bars), and 2 (purple error bars). The red dashed lines represent free particles energy level, solid red lines and red bands are center value and its error band of interacting energy levels;    (b)     Comparison of  two-particle energy spectra  lattice data  vs.  L\"uscher formula by using $G_0^{(a,L)} (0, E ) $  (red dashed curve), vs. L\"uscher formula result by using zero lattice spacing version of $G_0^{(L)} (0, E ) $ (blue dashed).
The model parameters are taken as:    $\kappa =0.1286$, and $\lambda  = 0.01$,    $T=120$,  $m=0.272$ and $V_0=0.165 \pm 0.04$. }\label{twomasslushcerplot}
\end{figure*}

  \begin{figure*}
 \centering
 \begin{subfigure}[b]{0.44\textwidth}
\includegraphics[width=0.99\textwidth]{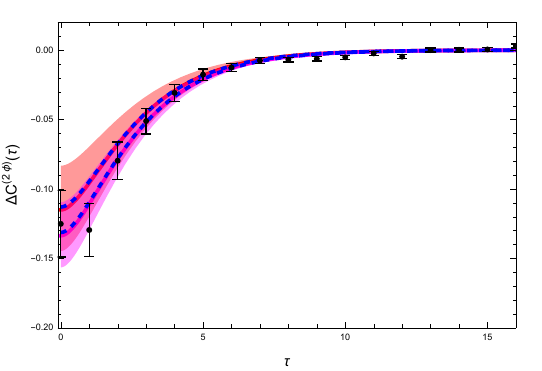}
\caption{  $\triangle C^{(2 \phi)} (  \tau) $ with two sets of $V_0$'s and fixed $L=40$.}\label{LuschertwodiffCtqplot}
\end{subfigure} 
\begin{subfigure}[b]{0.45\textwidth}
\includegraphics[width=0.99\textwidth]{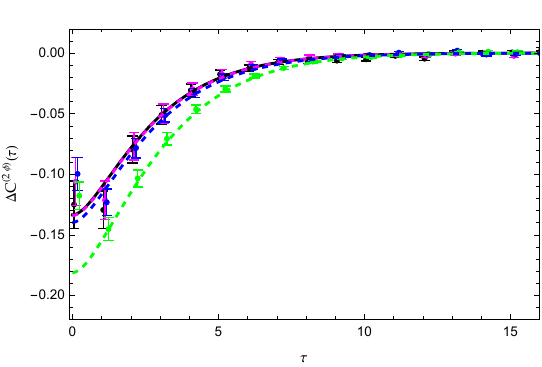}
\caption{   $\triangle C^{(2 \phi)} (  \tau) $ with various $L$'s and fixed $V_0 \sim 0.195$.}\label{sumdCtqplot}
\end{subfigure}
\caption{   (a) Comparison of    $\triangle C^{(2 \phi)} (  \tau) $  by using  spectral representation in Eq.(\ref{dCtspectralrep})  vs. lattice result (black error bars) for two sets of $V_0$'s:  $V_0=0.196 \pm 0.030$ (purple) and $V_0=0.165 \pm 0.040$ (red). The blue dashed curve is produced by using zero finite spacing limit version of  L\"uscher formula in Eq.(\ref{qccontactpot}); (b) Comparison of  spectral representation of   $\triangle C^{(2 \phi)} (  \tau) $ with various $L$'s for fixed  $V_0=0.196 \pm 0.030$: $L=10$(green), $20$(blue), $30$(purple) and $40$(black) vs. lattice data. Only center value curves are plotted.
The model parameters are taken as:    $T=120$ and  $m=0.272$. }\label{sumtwoCtplot}
\end{figure*}

\subsubsection{Comparison with  L\"uscher formula result}

The individual  two-particle energy levels can be extracted by applying the generalized eigenvalue method  \cite{Luscher:1990ck}, 
\begin{equation}
\widetilde{C}^{(2\phi)} (\tau) \xi_n = \lambda_n (\tau, \tau_0)   \widetilde{C}^{(2\phi)} (\tau_0) \xi_n,
\end{equation}
where $\tau_0$ is a small reference time and is set to zero in this work. The two-particle correlation function matrices is defined by
\begin{equation}
 \left [  \widetilde{C}^{(2\phi)}(\tau)   \right ]_{p, p'}  =  \sum_{x, x' \in [0, L-1]} e^{i p x} C^{(2\phi)}  (x \tau, x' 0) e^{- i p' x'}.
\end{equation}
A simple form of  $ \lambda_n (\tau,  0)  = e^{- E_n \tau}$ is used  in the data fitting for $\tau \in [0,8]$, see example of effective mass of two-particle in Fig.\ref{twomassplot}. To extract scattering phase shift or coupling strength for contact interaction,  the two-particle lattice  energy spectra  are fitted by using L\"uscher formula.

Two different version of quantization conditions are used in this work, first of all the zero lattice spacing limit version of quantization condition in Eq.(\ref{qccontactpot}) that is nothing but L\"uscher formula. Secondly, we also use  finite lattice spacing version of quantization condition by taking into consideration of  finite lattice spacing effect $(a=1)$,
\begin{equation}
\frac{1}{V_0} = G_{0}^{(a ,L)} (0, E ), \label{finitespacingQC}
\end{equation}
where   
\begin{equation}
G_{0}^{(a ,L)} (0, E ) = \frac{1}{L} \sum_{p = \frac{2\pi n}{L}}^{n \in  [ - \frac{L}{2}+1, \frac{L}{2}]} \frac{1}{ \omega_p^{(a,L)}   } \frac{1}{   E^2 - (2 \omega_p^{(a,L)} )^2  } .
\end{equation}
As described in Sec.~\ref{exactmodelcontactpot}, Eq.(\ref{finitespacingQC}) can be rewritten to the familiar  L\"uscher formula form.  
Two versions of quantization conditions make negligible  difference at low-lying energy spectra, see e.g. Fig.\ref{luscherplot}. The value of coupling strength extracted from lattice data is  $$V_0 = 0.165 \pm 0.040,$$ which is slightly lower than the value by fitting  the difference of integrated correlation functions.

\subsubsection{Spectral representation check}

As a consistent check,  we   also compute  the difference of integrated two particles correlation functions  by its spectral representation,
\begin{align}
   \triangle C^{(2\phi)}( \tau)  & \simeq    \sum_{p = \frac{2\pi n}{L}}^{n \in [ - \frac{L}{2} +1, \frac{L}{2}]}  \frac{\sigma_p }{2}   \left [    \frac{ e^{- E_p^{(a,L)}  \tau}  }{  E_p^{(a,L)}  } -     \frac{ e^{- 2 \omega_p^{(a,L)}  \tau}  }{  2 \omega_p^{(a,L)}  }    \right ] \nonumber \\
   & \stackrel[a \rightarrow 0 ]{T \rightarrow \infty}{  \rightarrow  }    \sum_{p = \frac{2\pi n}{L}}^{n \in [ 0,  \infty]}    \left [    \frac{ e^{- E_p  \tau}  }{  E_p   } -     \frac{ e^{- E^{(0)}_p   \tau}  }{  E^{(0)}_p   }    \right ] ,\label{dCtspectralrep}
\end{align}
where  the finite volume energy levels  are determined by    quantization condition in Eq.(\ref{finitespacingQC}) or L\"uscher formula at zero lattice spacing limit  in Eq.(\ref{qccontactpot}). In terms of spectral representation results,  we  observe that the lattice data seems prefer the value of coupling strength,  $V_0 = 0.196 \pm 0.030$,  which is extracted from integrated correlation function in Sec.\ref{V0bydCt}. The spectral representation of $ \triangle C^{(2\phi)}( \tau)$ with $V_0 \sim 0.196  $ agree well with lattice data. 
The example of comparison of $ \triangle C^{(2\phi)}( \tau)$  by spectral representation   vs. lattice data result is shown in Fig.~\ref{LuschertwodiffCtqplot} and Fig.~\ref{sumdCtqplot}, where the result  of    using zero lattice spacing limit version of L\"uscher formula in Eq.(\ref{qccontactpot}) is also plotted in Fig.~\ref{LuschertwodiffCtqplot}.    Again, the  finite lattice spacing quantization condition in Eq.(\ref{finitespacingQC})  and zero lattice spacing limit version of quantization condition in Eq.(\ref{qccontactpot}) makes negligible  difference in spectral representation results. The patten of convergence of both lattice data and spectral representation of $ \triangle C^{(2\phi)}( \tau)$  as increasing $L$  is displayed  in Fig.~\ref{sumdCtqplot}.

\subsubsection{A short summary}

With set of lattice model parameters:  $\kappa =0.1286$,   $\lambda  = 0.01$,  the difference of integrated correlation functions,  $ \triangle C^{(2\phi)}_{lat}( \tau)$,  for   various   $L$'s and a fixed $T=120$ are computed.  The mass of $\phi$ field is $m \sim 0.272$. The convergence of $ \triangle C^{(2\phi)}_{lat}( \tau)$ are observed and displayed in Fig.~\ref{twodiffCtqplot} and  Fig.~\ref{sumdCtqplot}.

The coupling strength of contact interaction potential is extracted by two different approaches: 
\begin{enumerate}[label=\arabic*)]
 \item  Fitting lattice data of $ \triangle C^{(2\phi)}_{lat}( \tau)$ with its infinite volume limit form  in right hand side of Eq.(\ref{mainresult}),  we find   the value of coupling strength:  $ V_0 =  0.196 \pm  0.03 0$, see Fig.~\ref{twodiffCtqplot}. 
 
  \item  The low-lying two-particle energy levels with various $L$'s ranging from $L=10$ up to $80$ are extracted from projected two-particle correlation functions, then fitting  lattice data of low-lying energy levels by using quantization condition in Eq.(\ref{finitespacingQC}),   the coupling strength is  thus extracted:   $ V_0 =  0.165 \pm  0.04 0$, see Fig.\ref{luscherplot}.
 
   \item The spectral representation of $ \triangle C^{(2\phi)}( \tau)$  in Eq.(\ref{dCtspectralrep}) are also computed with $V_0 = 0.196$ and plotted in Fig.~\ref{sumdCtqplot}. 
\end{enumerate}

Overall, the result by using L\"uscher formula to fit low-lying energy spectra and the result  by fitting $ \triangle C^{(2\phi)}( \tau)$ agree with each other within errors. The possible cause on the difference of  $V_0$'s from two approaches may  be that L\"uscher formula fit only low-lying energy spectra, however integrated correlation function  is the result of sum of all energy levels.   Fitting the difference of integrated  correlation functions is  more like a "global fit".  In principle, the lattice data in integrated correlation function contain inelastic contribution as well, however, the main result in Eq.(\ref{mainresult}) is only formulated based on the assumption of  existence  of elastic channel only. The inelastic contributions are suppressed by exponentially decaying factor $ \frac{ e^{-E_n t} }{E_n}$ for large $\tau$. However, since  the mass of single $\phi$ field is only $m\sim 0.272$, four particles threshold start at $4 m \sim 1.1$ which is not too heavy,  they may still have some residual effects near $\tau \sim 0$. The data in Fig.~\ref{twodiffCtqplot} and Fig.~\ref{sumdCtqplot} near $\tau \sim 0$ start going wild, which may be the indication of inelastic contribution.

  \begin{figure*}
 \centering
 \begin{subfigure}[b]{0.45\textwidth}
\includegraphics[width=0.99\textwidth]{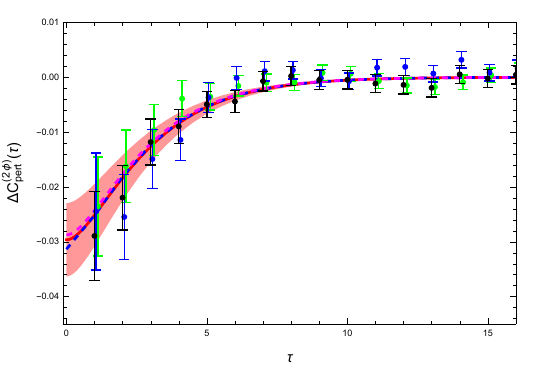}
\caption{ Perturbation result  with $V_0 \sim 0.025 $  vs. lattice data.}\label{LusPertdiffCtqplot}
\end{subfigure} 
\begin{subfigure}[b]{0.43\textwidth}
\includegraphics[width=0.99\textwidth]{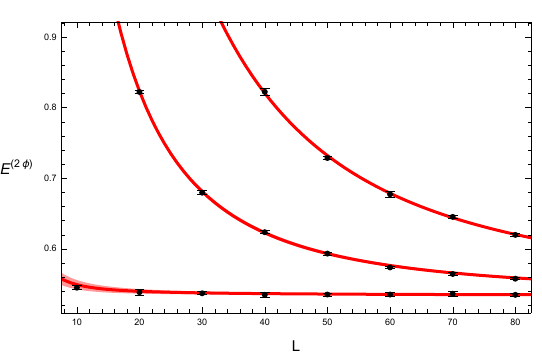}
\caption{  $  E^{(2\phi)}  $ with weak coupling $V_0 \sim 0.025 $.}\label{twomassallLpertplot}
\end{subfigure}
\caption{   (a) Comparison of  complete perturbation result of $\triangle C^{(2 \phi)}_{pert} (  \tau) $ given  in Eq.(\ref{dCpertfinitelatticespacing}) (dashed blue) vs. approximation expression in Eq.(\ref{dCpertapprox}) (red band) vs. $\triangle C^{(2 \phi)} (  \tau) $ by spectral representation in Eq.(\ref{dCtspectralrep}) (dashed purple) vs. lattice data of $L=30$(green), $40$(blue) and $50$(black error bars);
    (b)     Comparison of  two-particle energy spectra  lattice data  vs.  spectral generated by L\"uscher formula     (red band)  for weak coupling $V_0 \sim 0.025 $.
The model parameters are taken as:    $\kappa =0.1235$,   $\lambda  = 0.001$,    $T=120$ and  $m=0.267$. }\label{twoCtPertplot}
\end{figure*}

  \begin{figure*}
 \centering
 \begin{subfigure}[b]{0.45\textwidth}
\includegraphics[width=0.99\textwidth]{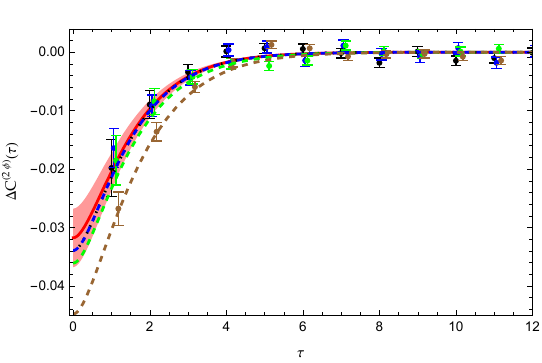}
\caption{ Heavy $\phi$ field  with $m \sim 0.500$   vs. lattice data.}\label{heavysumdCtqplot}
\end{subfigure} 
\begin{subfigure}[b]{0.43\textwidth}
\includegraphics[width=0.99\textwidth]{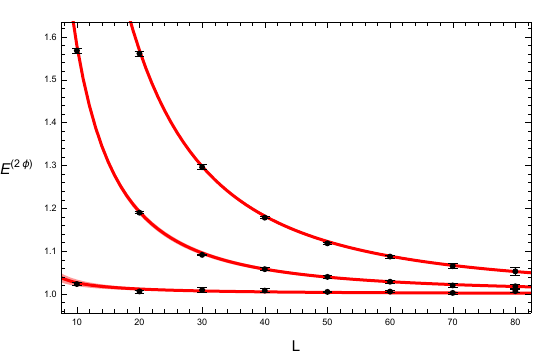}
\caption{  $  E^{(2\phi)}  $ with $m \sim 0.500$ and  $V_0 \sim 0.271 $}\label{heavytwomassallLplot}
\end{subfigure}
\caption{   (a) Comparison of  heavy $\phi$ field infinite volume limit   of $\triangle C^{(2 \phi)}  (  \tau) $   (solid red band) vs.   spectral representation of $\triangle C^{(2 \phi)}  (  \tau) $   in Eq.(\ref{dCtspectralrep}) (dashed colored) vs. lattice data of $L=6$(brown), $L=10$(green), $20$(blue) and $30$(black error bars);
    (b)     Comparison of  two-particle energy spectra  lattice data  vs.  spectral generated by L\"uscher formula     (red dashed curve)  for heavy $\phi$ particle and coupling $V_0 \sim 0.271 $.
The model parameters are taken as:    $\kappa =0.122$,   $\lambda  = 0.01$,    $T=120$ and  $m=0.500$. }\label{twoCtHeavyplot}
\end{figure*}

We remark that   for the  simple $1+1$ dimensional  lattice model with a contact or short-range interaction potential,  the scattering phase shift   can be parameterized by a single parameter, $V_0$, which represents the strength of short-range potential:
\begin{equation}
\delta(E) = \cot^{-1} \left [  \frac{ Re \left [ G_0^{(\infty)} (0; E)  \right ] - \frac{1}{V_0} }{\rho(E)}  \right ],
\end{equation} 
where analytic expression of $G_0^{(\infty)} (0; E) $ is given in Eq.(\ref{inffreeGreenanaly}). In $1+1$ dimensions, contact interaction potential strength, $V_0$, is free off ultra-violet divergence, so it is convenient to use it as a free parameter for scattering phase shift directly at the scope of current discussion. In general, the scattering phase shift is usually parametrized in terms of a few free parameters and kinematic factors  based on either chiral perturbation theory or $K$-matrix formalism,  see e.g. \cite{Guo:2013vsa,Guo:2012hv}. The free parameters of analytic form of scattering phase shift can be associated with the renormalized coupling strength, mass of resonance, etc. For instance, the coupling strength of contact interaction potential in $3+1$ dimensions suffers ultra-violet divergence and must be renormalized,   the parameterization of scattering phase shift is hence given in terms of the renormalized coupling strength, see e.g. Eq.(75) in \cite{Guo:2020kph}.

\subsection{Sanity check on inelastic contribution}
As another sanity  check about consistency between L\"uscher formula and fitting $\triangle C^{(2 \phi)} (  \tau) $ approach, two possible ways of suppressing inelastic channel contribution may be: (1) reducing coupling strength, since four-particle contribution may start at the order of $V^2_0$; and (2) increasing the mass of $\phi$ field, hence the threshold of four-particle is lifted and inelastic contribution is suppressed by $\frac{e^{-E_n \tau}}{E_n}$ factors.

\subsubsection{Perturbation calculation check}
 First of all we   reduce coupling strength and compare results from lowest order perturbation calculation,  spectral representation of  $\triangle C^{(2 \phi)} (  \tau) $   in Eq.(\ref{dCtspectralrep}), and  also low-lying two-particle spectra. The lattice model parameters are reset to $\kappa =0.1235$ and $\lambda  = 0.001$. The single particle mass  and  coupling strength now are around $m  =  0.267 \pm  0.002   $  and  $V_0 = 0.025 \pm 0.008$ respectively.

The perturbation calculation in Euclidean spacetime can be carried out as described in Sec.~\ref{perturbationresult}, also taking into consideration of finite lattice spacing, we find
\begin{equation}
\triangle C^{(2 \phi)}_{pert} (  \tau) =   \frac{1}{L} \sum_{p = \frac{2\pi n}{L}}^{n \in    [- \frac{L}{2} +1, \frac{L}{2}] }  2 \omega^{(a, L)}_p  \triangle \widetilde{C}_{pert}^{(2\phi)}(p, \tau)  , \label{dCpertfinitelatticespacing}
\end{equation}
where
\begin{equation}
\triangle \widetilde{C}^{(2 \phi)}_{pert} ( p, \tau) = - \frac{V_0}{T}  \sum_{\omega = \frac{2\pi n}{L}}^{n \in [0, T-1]}  e^{i \omega \tau} \left [ G_{2\phi}  (p, \omega) \right ]^2  ,
\end{equation}
and the finite volume two-particle Green's function is defined by
\begin{align}
& G_{2\phi}  (p, \omega)  \nonumber \\
&= \frac{1}{T}  \sum_{\omega' = \frac{2\pi n'}{L}}^{n' \in [0, T-1]}   \frac{1}{2 -2  \cos \omega'     - 2   \cos   k   + 2 \cosh m  } \nonumber \\
& \times  \frac{1}{2 -2  \cos (\omega - \omega' )    - 2   \cos   k   + 2 \cosh m  } .
\end{align}
At    the  limit of $T \rightarrow \infty$ and   zero  lattice spacing,  
\begin{equation}
  G_{2\phi}  (p, \omega)   \stackrel[a\rightarrow 0]{T \rightarrow \infty}{ \rightarrow }   \frac{1}{\omega_p} \frac{1}{\omega^2+ (2\omega_p)^2}    ,
\end{equation}
and
\begin{equation}
\triangle \widetilde{C}^{(2 \phi)}_{pert} ( p, \tau)   \stackrel[a\rightarrow 0]{T \rightarrow \infty}{ \rightarrow }  - V_0  \frac{\tau + \frac{1}{2\omega_p}}{(2\omega_p)^4} e^{- 2 \omega_p \tau} .
\end{equation}
Hence, the perturbation result at the limit of zero lattice spacing in Eq.(\ref{perturbfinitevolume})  is recovered.  It can be shown numerically, see e.g. Fig.\ref{LusPertdiffCtqplot}, that  except  at $\tau=0$,  the Eq.(\ref{dCpertfinitelatticespacing}) can be well approximated by
\begin{equation}
\triangle C^{(2 \phi)}_{pert} (  \tau)  \simeq  -  \frac{V_0}{L} \sum_{p = \frac{2\pi n}{L}}^{n \in  [- \frac{L}{2} +1, \frac{L}{2}] }   \frac{\tau + \frac{1}{2 \omega^{(a,L)}_p}}{  ( 2 \omega^{(a,L)}_p )^3} e^{- 2 \omega^{(a,L)}_p \tau}  .  \label{dCpertapprox}
\end{equation}

The example of lattice data of $\triangle C^{(2 \phi)} (  \tau) $ vs. perturbation results is plotted in Fig.\ref{LusPertdiffCtqplot}, as the comparison, the result of $\triangle C^{(2 \phi)} (  \tau) $  by computing spectral representation in Eq.(\ref{dCtspectralrep}) is also shown in  Fig.\ref{LusPertdiffCtqplot}. With  the same set of coupling strength,  $V_0 = 0.025 \pm 0.008$, the low-lying two-particle spectral are computed by using   finite lattice spacing version    quantization condition in  Eq.(\ref{finitespacingQC}) compared with lattice data, see Fig.\ref{twomassallLpertplot}. Overall, the consistency is excellent.

\subsubsection{Heavy \texorpdfstring{$\phi$}{phi} field check}

Next we also increase the mass of $\phi$ field by setting lattice model parameters to $\kappa =0.122$ and $\lambda =0.01$. The single particle mass and coupling strength in this case are around $m = 0.500 \pm 0.001$ and $V_0 = 0.271\pm 0.030$ respectively.

The   lattice data of $\triangle C^{(2 \phi)} (  \tau) $ vs.  its infinite volume limit results are plotted in Fig.\ref{heavysumdCtqplot},   the result of $\triangle C^{(2 \phi)} (  \tau) $  by computing spectral representation in Eq.(\ref{dCtspectralrep}) is also shown in  Fig.\ref{heavysumdCtqplot}.  With the same set of coupling strength,  $V_0 = 0.271 \pm 0.030$, the low-lying two-particle spectral are computed by using   finite lattice spacing version    quantization condition in  Eq.(\ref{finitespacingQC}) compared with lattice data, see Fig.\ref{heavytwomassallLplot}.   Again,  both approaches show excellent consistency.

\section{Summary and discussion}\label{summary}

In summary, a relativistic formalism that connects the difference of interacting and non-interacting two-particle correlation functions  to the scattering phase shift by an integral is derived, the main result is shown in Eq.(\ref{mainresult}).  The difference of finite volume  two-particle correlation functions converges rapidly to its infinite volume limit near small Euclidean times as the size of finite box is increased. Hence the proposed approach may have a good potential to overcome S/N problem in lattice calculation of two-nucleon interactions.

The numerical tests are conducted by (1) analytic solutions of a exactly solvable contact interaction model, (2) perturbation calculation, and (3) Monte Carlo simulation of $\phi^4$ lattice field theory model.

\begin{table}[htp]
\caption{List of sets of $\phi^4$ model parameters $(\kappa, \lambda)$, and corresponding      single particle mass and  coupling strength $V_0$.}
\begin{center}
\begin{tabular}{|cc|c|c|}
\hline
$(\kappa, $ &  $ \lambda) $ & $m$ & $V_0$ \\
\hline
(0.1213, & 0  )& $ 0.350\pm 0.003$ & $0$  \\
(0.1286, & 0.01  )& $ 0.272\pm 0.003$ & $0.20 \pm 0.03$ vs. $0.17 \pm 0.04$ \\
(0.1235, & 0.001  )& $ 0.267\pm 0.003$ & $ 0.025 \pm 0.008$ \\
(0.1220, & 0.01  )& $ 0.500\pm 0.001$ & $ 0.27 \pm 0.03$ \\
\hline
\end{tabular}
\end{center}
\label{paramtable}
\end{table}%

  In Monte Carlo simulation of $\phi^4$ theory, the model with four different sets of parameters are calculated, the single particle mass and coupling strength are extracted accordingly, see Table \ref{paramtable}. The first set $(\kappa, \lambda) = (0.1213, 0)$ represents non-interacting $\phi$ fields, and it is used to check lattice dispersion relation, and agreements of analytic expression of non-interacting correlation functions vs. lattice data. The second set $(\kappa, \lambda) = (0.1286, 0.01)$ represents interacting particles with mass of $m \sim 0.272$ and coupling strength of $V_0 \sim  0. 20 $ (our approach) or $V_0 \sim 0.17$ (L\"uscher formula approach). The coupling strength extracted from two different approach differ slightly within errors. The difference may be caused inelastic channel contribution: L\"uscher formula approach fit only low-lying energy spectra, but integrated correlation function approach fit the sum of all energy levels and may be considered as a "global fit" approach. The inelastic contributions  are  not yet included  in  our formalism at current scope, which may show at small Euclidean time region.   At current  $\phi^4$ lattice model, there are no three-to-two particles coupling, so the inelastic threshold starts at $ 4 m \sim 1.1 $, the inelastic effects are not significantly suppressed by $\frac{e^{-E_n \tau}}{E_n}$ near $\tau \sim 0$.  As a sanity check, two other sets of parameters are chosen to simulate scenarios of (1) weak coupling strength $V_0 \sim 0.025$ with $m \sim 0.267$, and (2) heavy mass $m \sim 0.5$ with $V_0 \sim 0.27$. The inelastic contributions are suppressed in both scenarios: (1)  four-particle interaction show up at order of $V_0^2$ in weak coupling scenario, and (2) $4\phi$ threshold starts now at $4 m\sim 2$ in heavy mass scenario. Numerically, both scenarios show good agreement between proposed approach and L\"uscher formula approach.  On the other hand,  the inelastic effect and  coupled-channel dynamics should be installed and studied  in a more rigorous way, which will be carried out in our further publications.

At last, we would like to comment that both L\"uscher formula approach and the integrated correlation function approach proposed in this manuscript have their own pros and cons. L\"uscher formula offers a   model-independent way of extracting elastic scattering phase shifts, which convert one energy level from lattice calculation into one point of phase shift   directly. However it suffers   difficulties at large volume calculation or S/N problem. On the other hand, the integrated correlation function approach  in principle is also a model independent approach, see e.g. Eq.(\ref{mainresult}). However the difference of integrated correlation function is related to the phase shift by an integral, hence the phase shift cannot be pulled out from lattice data  point-by-point directly. Instead, we have to firstly model and parametrize the scattering phase shift in terms of a few parameters and kinematic factors based on either chiral perturbation theory or $K$-matrix formalism, etc. The phase shift can then be obtained by fitting lattice data of integrated correlation functions with these free parameters through an integral. The advantage of  the integrated correlation function approach is that it shows the rapid convergence as the volume is increased and it also show  potentials to overcome S/N problem, which may be useful in nucleon-nucleon interaction lattice calculation.

\acknowledgments
We   acknowledge support from  the College of Arts and Sciences and Faculty Research Initiative Program,  Dakota State University, Madison, SD.  P.G. acknowledges   computing resources (https://cloud.madren.org/)  at DSU made available for conducting the research reported in this work. This research was supported in part by the National Science Foundation under Grant No. NSF PHY-1748958.

\appendix

\section{Relativistic two-particle scattering solutions in infinite volume}\label{scattsolutionsinfvol}

\subsection{Relativistic Lippmann-Schwinger like equation and scattering amplitude}
For the short-range interaction, the  scattering of relativistic particles may be well described by relativistic Lippmann-Schwinger like equation, see e.g. Refs.~\cite{Guo:2019ogp,Guo:2020kph}, where the relativistic   Lippmann-Schwinger like equations can be derived from Bethe-Salpeter equations with assumption of "instantaneous interaction kernel"  \cite{Guo:2019ogp,Guo:2020kph}.

In present work, a contact interaction is considered, $$V(r) = V_0 \delta(r),$$ hence only even parity states are affected by interaction. The scattering solutions can be  found in Appendix B in Ref.~\cite{Guo:2020kph}. The relative scattering wave function satisfies LS like equation,
\begin{equation}
\psi^{(\infty)}_{E_k}  (r)= \cos (k r) + G^{(\infty)}_0(r, E_k) V_0 \psi^{(\infty)}_{E_k}  (0),
\end{equation}
where $k = \sqrt{\frac{E^2_k}{4} - m^2}$ is the relative momentum of two particles in CM frame, and free particles Green's function is defined by
\begin{align}
G^{(\infty)}_0 (r; E)  =  \int_{-\infty}^\infty \frac{d q}{2\pi}  \frac{1}{ \sqrt{q^2+ m^2}} \frac{e^{i q r} }{E^2 -  4 (q^2+ m^2) + i 0 }. \label{infinitevolumefreeGreen}
\end{align}

The scattering amplitude is introduced by
\begin{equation}
t(E) = - \frac{1}{\frac{1}{V_0} - G^{(\infty)}_0 (0; E) }, \label{scattamp}
\end{equation}
hence the   wave function can be rewritten as
\begin{equation}
\psi^{(\infty)}_{E_k}  (r)= \cos (k r) -  t(E_k) G^{(\infty)}_0(r, E_k)  .
\end{equation}
The  function  $G^{(\infty)}_0 (0; E)$  is an analytic function of $E^2$ with a branch cut starting at threshold $4m^2$,
   \begin{equation} 
G^{(\infty)}_0 (0; E)   =  \frac{ 1 }{2\pi} \int_{4 m^2}^\infty d s'   \frac{ 1  }{ \sqrt{s' (   s' - 4 m^2 ) }  }  \frac{ 1  }{E^2 -  s' + i 0 }.
\end{equation}
The analytic expression of $G^{(\infty)}_0 (0; E)$ can be found rather straightforwardly,
\begin{align}
&G^{(\infty)}_0 (0; E) = \frac{\rho(E) }{\pi}    \nonumber \\
&  \times \left [   \ln  \frac{E^2-2 m^2 +\sqrt{E^2(E^2 - 4 m^2)} }{2 m^2} - i  \pi \theta( E - 2 m) \right ], \label{inffreeGreenanaly}
\end{align}
where
\begin{align}
  \rho(E) =  \frac{1}{2} \frac{1}{\sqrt{E^2(E^2 - 4 m^2)}}. \label{infphasefactor}
\end{align}
The scattering amplitude can be parameterized by a phase shift,
\begin{equation}
t(E)  = \frac{1}{\rho(E)} \frac{e^{2 i \delta(E)} -1}{2i} = \frac{1}{\rho(E)} \frac{1}{\cot \delta(E) - i }, \label{scattampparametrization}
\end{equation}
and the on-shell unitarity relation is determined by,
\begin{equation}
Im  \left [ t^{-1}(E) \right ] = -   \theta( E - 2 m) \rho(E). \label{unitarityrelation}
\end{equation}

The scattering amplitude can also be expressed in terms of  Muskhelishvili-Omn\`es (MO) representation \cite{muskhelishvili1941application,Omnes:1958hv}  that is sometimes also referred to $N/D$ method \cite{PhysRev.119.467,PhysRev.130.478},   also see e.g. Refs.~\cite{Gorchtein:2011vf,Danilkin:2014cra,Guo:2014vya,Guo:2015zqa,Guo:2016wsi,Guo:2022row}, 
\begin{align}
t(E)   = N e^{ \frac{1}{\pi} \int_{4 m^2}^\infty d s \frac{ \delta(\sqrt{s}) }{ s - E^2 - i 0} } , \label{MOscattamp}
\end{align}
where $N= t(2m)e^{ - \frac{1}{\pi} \int_{4 m^2}^\infty d s \frac{ \delta(\sqrt{s}) }{ s - 4m^2 - i 0} } $ is a constant factor.

\subsection{Two particles Green's function and its spectral representation}

The relativistic two-particle Green's function may be introduced by Dyson equation,
\begin{align}
& G^{(\infty)} (r,r' ; E)  \nonumber \\
& =  G^{(\infty)}_0 (r-r'; E)  + G^{(\infty)}_0 (r; E)  V_0 G^{(\infty)} (0,r' ; E),
\end{align}
the analytic solution is thus given by
\begin{align}
& G^{(\infty)} (r,r' ; E) - G^{(\infty)}_0 (r-r'; E)   \nonumber \\
&  =  - G^{(\infty)}_0 (r; E) t(E)   G_0^{(\infty)} (r' ; E). \label{infinitevolumeGreendiff}
\end{align}
The spectral representation of Green's function is given by
\begin{align}
& G^{(\infty)} (r, r' ; E)  \nonumber \\
& =  \int_{-\infty}^\infty \frac{d q}{2\pi}  \frac{1}{ \sqrt{q^2+ m^2}} \frac{ \psi^{(\infty)}_{E_q}  (r) \psi^{(\infty)*}_{E_q}  (r') + \sin (q r) \sin (q r') }{E^2 -  E_q^2 + i 0 },
\end{align}
where $E_q = 2 \sqrt{q^2+ m^2}$. The $ \sin (q r) $ is  wave function of odd parity states that will be cancelled out between interacting and non-interacting Green's function and do not contribute on the right hand side of Eq.(\ref{infinitevolumeGreendiff}).

\subsection{Integrated Green's function and its relation to scattering phase shift}\label{relatFriedelformula}
Using Eq.(\ref{infinitevolumeGreendiff}) and analytic expression of free particles Green's function in Eq.(\ref{infinitevolumefreeGreen}), we find 
\begin{align}
& \widetilde{G}^{(\infty)} (p,p ; E) - \widetilde{G}^{(\infty)}_0 (p-p; E)   \nonumber \\
&  =    \frac{1}{ \sqrt{p^2+ m^2}}    t(E)    \frac{d}{d E^2} \left [  \frac{1}{ \sqrt{p^2+ m^2}} \frac{1 }{E^2 - E_p^2  + i 0 }   \right ]   ,
\end{align}
where the momentum space Green's function is defined by
\begin{equation}
\widetilde{G}^{(\infty)} (p,p' ; E)  = \int_{-\infty}^\infty d r d r'  e^{i p r} G^{(\infty)} (r, r' ; E)  e^{- i p' r' }.
\end{equation}
Hence the difference of  integrated Green's functions is given in terms of scattering amplitude by
\begin{align}
& \int_{-\infty}^\infty \frac{d p}{2\pi} \sqrt{p^2+ m^2} \left [ \widetilde{G}^{(\infty)} (p,p ; E) - \widetilde{G}^{(\infty)}_0 (p-p; E)  \right ]  \nonumber \\
&  =    \frac{d}{d E^2}  \ln \left [   t(E)  \right ] .
\end{align}
Using MO representation of scattering amplitude in Eq.(\ref{MOscattamp}), thus we finally find 
\begin{align}
& \int_{-\infty}^\infty \frac{d p}{2\pi} \sqrt{p^2+ m^2} \left [ \widetilde{G}^{(\infty)} (p,p ; E) - \widetilde{G}^{(\infty)}_0 (p-p; E)  \right ]  \nonumber \\
&  =    -   \frac{1}{\pi} \int_{4 m^2}^\infty d s \frac{ \delta(\sqrt{s}) }{ (s - E^2 - i 0)^2 }   . \label{Friedelformula}
\end{align}

The imaginary part of Eq.(\ref{Friedelformula}) yields
\begin{align}
&  \int_{-\infty}^\infty \frac{d p}{2\pi}  \frac{1}{2  \omega_p }  \left [   \widetilde{ \psi}^{(\infty)}_{E_k}  (p)  \widetilde{\psi}^{(\infty)*}_{E_k}  (p)   -   \widetilde{ \psi}^{(0, \infty)}_{E_k}  (p)  \widetilde{\psi}^{(0,\infty)*}_{E_k}  (p)  \right ]   \nonumber \\
&    =   - 4 k    \frac{d}{d E_k}      \delta( E_k )   , \label{normalizationinfwav}
\end{align}
where
\begin{equation}
 \widetilde{ \psi}^{(0, \infty)}_{E_k}  (p) = 2 \omega_p \frac{\delta_{k, p} + \delta_{k, - p}}{2} .
\end{equation}
The Eq.(\ref{normalizationinfwav}) may be considered as normalization of scattering wave function, the finite term on the right hand side of equation is the result of boundary condition of scattering solutions, see discussion in Ref.\cite{Poliatzky:1992gn}. As the matter of fact, when condition
\begin{align}
  \int_{-\infty}^\infty \frac{d p}{2\pi}  \frac{1}{2  \omega_p }    \widetilde{ \psi}^{(\infty)}_{E}  (p)  \widetilde{\psi}^{(\infty)*}_{E'}  (p)  = 0 \ \ for \ \ E \neq E'
\end{align}
 is imposed,   the on-shell unitarity relation of scattering amplitude in Eq.(\ref{unitarityrelation}) can be further generalized to
 \begin{align}
t(E) - t^*(E') = \left [G_0^{(\infty)} ( 0; E)  - G_0^{(\infty) *} ( 0; E')  \right ]t^*(E') t(E).
\end{align}

\section{Relativistic Lippmann-Schwinger like equation in finite volume and random phase approximation}\label{RPAappendix}

In this section, we show that  the relativistic     Lippmann-Schwinger like equations can be derived from random phase approximation (RPA) \cite{Fetter}. The  Hamiltonian of the lattice field theory defined  in Eq.(\ref{Minkowskiaction}) is given by the free particles term
\begin{equation}
\hat{H}_0 = \frac{1}{L} \sum_{p = \frac{2\pi n}{L}}^{n \in \mathbb{Z}} \omega_p  ( \hat{a}_p^\dag \hat{a}_p + \hat{b}_p^\dag \hat{b}_p   ),
\end{equation}
and  the interaction term,
\begin{equation}
\hat{H}_I  =    \frac{1}{4 !}    \int_{0}^L d x d y | \hat{ \phi }(x)   |^2V(x-y) | \hat{ \phi }(y)   |^2,  
\end{equation}
where
\begin{equation}
 \hat{\phi }(x) = \frac{1}{L}\sum_{p = \frac{2\pi n}{L}}^{n \in \mathbb{Z}} \frac{1}{\sqrt{2 \omega_p}} \left [  e^{ i p x} \hat{a}_p^\dag + e^{- i p x} \hat{b}_p    \right ],
\end{equation}
where $\omega_p = \sqrt{p^2+m^2}$.
The $\hat{a}_p^\dag (\hat{a}_p)$ and $\hat{b}_p^\dag (\hat{b}_p)$ are creation (annihilation) operators of particles and antiparticles respectively, they satisfies commutation relations,
\begin{equation}
\left [ \hat{a}_p,   \hat{a}_{p'}^\dag  \right ] = \left [ \hat{b}_p,   \hat{b}_{p'}^\dag  \right ]  = L  \delta_{p, p'}.
\end{equation}

Based on RPA assumption,  the two particles states can be created by creating two particles from vacuum or equivalently by annihilate two antiparticles in the vacuum, so the two particles wave function is defined by
\begin{equation}
\widetilde{\psi}^{(a)}_E (p) = \langle E |  \frac{1}{\sqrt{2}}   \hat{a}_{p}^\dag   \hat{a}_{-p}^\dag  | 0 \rangle  ,
\end{equation}
 and
 \begin{equation}
\widetilde{\psi}^{(b)}_E (p) =   \langle E |  \frac{1}{\sqrt{2}}   \hat{b}_{p}   \hat{b}_{-p}  | 0 \rangle ,
\end{equation}
where superscripts  $(a,b)$  are used to label the two-particle wave functions created by two different mechanisms.

The dynamical equations of two-particle states under RPA are derived from
\begin{equation}
\langle E | \left [  \hat{H} ,  \frac{1}{\sqrt{2}}   \hat{a}_{p}^\dag   \hat{a}_{-p}^\dag \right ]  | 0 \rangle = E \widetilde{\psi}^{(a)}_E (p) ,
\end{equation}
and
\begin{equation}
\langle E | \left [  \hat{H} ,  \frac{1}{\sqrt{2}}   \hat{b}_{p}   \hat{b}_{-p} \right ]  | 0 \rangle = E \widetilde{\psi}^{(b)}_E (p) ,
\end{equation}
where $ \hat{H}  =  \hat{H}_0  +  \hat{H}_I $ is full Hamiltonian operator. After some length calculation, we finally find  coupled equations,
\begin{align}
 & \widetilde{\psi}^{(a)}_E (p)  = \frac{1}{2 \omega_p } \frac{1}{E - 2 \omega_p }  \nonumber \\
& \times \frac{1}{L} \sum_{p' = \frac{2\pi n' }{L}}^{n' \in \mathbb{Z}} \frac{1}{2 \omega_{p'}} \widetilde{V} (p - p') \left [\widetilde{\psi}^{(a)}_E (p')   + \widetilde{\psi}^{(b)}_E (p')   \right ],
\end{align}
and
\begin{align}
 & \widetilde{\psi}^{(b)}_E (p)  = - \frac{1}{2 \omega_p } \frac{1}{ E +  2 \omega_p }  \nonumber \\
& \times \frac{1}{L} \sum_{p' = \frac{2\pi n' }{L}}^{n' \in \mathbb{Z}} \frac{1}{2 \omega_{p'}} \widetilde{V} (p - p') \left [\widetilde{\psi}^{(a)}_E (p')   + \widetilde{\psi}^{(b)}_E (p')   \right ],
\end{align}
where $\widetilde{V} (p) = \int_0^L d r e^{i p r} V(r)$ is Fourier transform of interaction potential.  Two equations can be combined together by defining 
\begin{equation}
 \widetilde{\psi}_E (p) = \widetilde{\psi}^{(a)}_E (p) +  \widetilde{\psi}^{(b)}_E (p) ,
\end{equation}
we thus find
\begin{equation}
  \frac{ \widetilde{\psi}_E (p)  }{2\omega_p} = \frac{1}{ \omega_p } \frac{1}{E^2 - (2 \omega_p)^2 }   \frac{1}{L} \sum_{p' = \frac{2\pi n' }{L}}^{n' \in \mathbb{Z}} \widetilde{V} (p - p')  \frac{\widetilde{\psi}_E (p')   }{2 \omega_{p'}} .
\end{equation}
Next let's define coordinate space wave function by
\begin{equation}
 \psi^{(L)}_E (r) =  \frac{1}{L} \sum_{p = \frac{2\pi n }{L}}^{n \in \mathbb{Z}}  e^{- i p r} \frac{ \widetilde{\psi}_E (p)  }{2\omega_p} ,
\end{equation}
we thus find
\begin{equation}
 \psi^{(L)}_E (r) =  \int_0^L d r \left [ \frac{1}{L} \sum_{p = \frac{2\pi n }{L}}^{n \in \mathbb{Z}}   \frac{1}{ \omega_p } \frac{e^{i p (r- r')}}{E^2 - (2 \omega_p)^2 } \right ] V(r')  \psi^{(L)}_E (r') ,
\end{equation}
which is consistent with result that is derived from Bethe-Salpeter equations with assumption of "instantaneous interaction kernel"  \cite{Guo:2019ogp,Guo:2020kph}.

\bibliography{ALL-REF.bib}

\end{document}